\documentclass{article}

\usepackage[]{arxiv}
\usepackage{setspace}
\usepackage{float}
\usepackage{amsmath}
\usepackage{multirow}
\usepackage[utf8]{inputenc} 
\usepackage[T1]{fontenc}    
\usepackage[colorlinks]{hyperref}       
\usepackage{url}            
\usepackage{booktabs}       
\usepackage{amsfonts}       
\usepackage{nicefrac}       
\usepackage{microtype}      
\usepackage{lipsum}
\usepackage{graphicx}
\usepackage[table]{xcolor}
\usepackage{braket}
\graphicspath{ {./images/} }
\usepackage{caption}
\usepackage{subcaption}
\usepackage{biblatex}
\usepackage{cleveref}
\usepackage{amssymb}
\usepackage{stackengine} 
\usepackage{comment}
\usepackage{lineno}

\title{BOLDreams: Dreaming with pruned in-silico fMRI Encoding Models of the Visual Cortex}
\author{Uzair Hussain \thanks{Corresponding author}\\ Krembil Research
Institute, University Health Network, Toronto, Canada \\
\texttt{ughussain@gmail.com} \\
\And  Kamil Uludag \\
Department of Medical Biophysics, University of Toronto, Toronto, Ontario,
Canada \\
Techna Institute \& Koerner Scientist in MR Imaging, University Health Network,
Toronto, Canada\\
Center for Neuroscience Imaging Research, Sungkyunkwan University, Suwon,
Republic of Korea\\
Physical Sciences, Sunnybrook Research Institute, Toronto, ON, Canada\\
\texttt{kamil.uludag@uhn.ca}
}

\begin{document}
\let\ref\autoref
\maketitle
\begin{abstract}
In this article we use the Natural Scenes Dataset (NSD) to train a family of
feature-weighted receptive field neural encoding models. These models use a
pre-trained vision or text backbone and map extracted features to the voxel
space via receptive field readouts. We comprehensively assess such models,
quantifying performance changes based on using different modalities like text or
images, toggling finetuning, using different pre-trained backbones, and changing
the width of the readout. We also dissect each model using explainable AI (XAI)
techniques, such as feature visualization via input optimization, also referred
to as ``dreaming'' in the AI literature, and the integrated gradients approach
to calculate implicit attention maps to illustrate which features drive the
predicted signal in different brain areas. These XAI tools illustrate
biologically plausible features that drive the predicted signal. Traversing the
model hyperparameter space reveals the existence of a maximally minimal model,
balancing simplicity while maintaining performance.
\end{abstract}



\section{Introduction}
Convolution neural networks (CNNs) are one of the most commonly used
architectures in computer vision problems. First introduced in the late 90’s by
LeCunn et. al. \cite{lecun_gradient-based_1998} for document recognition, they
gained renowned popularity with the success of AlexNet
\cite{krizhevsky_imagenet_2012} for winning the ImageNet classification
challenge with a substantial margin. In a broad sense, CNNs can be viewed as
hierarchical feature extractors; in shallow layers fine-grained details are
extracted which are then combined in deeper layers to make complex image
features. This architectural choice is inspired by models of the human visual
cortex \cite{agrawal_convolutional_nodate}. In fact, in early layers of the CNN,
filters emerge that detect orientation of edges at various angles (Gabor
filters), similar to orientation sensitive neurons in the early visual cortex
\cite{hubel_sequence_1974}. This makes CNNs a good candidate for modeling
neuronal activity in the visual cortex, although these are often considered to be “black box” models,
since CNNs are quite complex and difficult to interpret. Many explainable AI
(XAI) techniques have emerged that aim to alleviate this opaqueness of CNNs. One
theme of our work is to translate these techniques to understand models of the
visual cortex that are based on CNNs. 

Naturally, to build such CNN based models of the visual cortex we need a probe
into neuronal activity. Direct in-vivo recording of neurons poses significant
challenges, therefore, a popular alternative is functional magnetic resonance
imaging (fMRI). A commonly used contrast in fMRI imaging is the Blood
Oxygenation Level Dependent (BOLD) contrast. This contrast is based on a
remarkably convenient property of blood; oxygenated hemoglobin (Hb) is
diamagnetic, whereas deoxygenated hemoglobin is paramagnetic and has a higher
magnetic susceptibility, thus causing a drop in the signal
\cite{huettel_functional_2014}. Herein lies a shortcoming of this approach, BOLD
does not measure neuronal activity directly but rather is a measure of the
metabolic demands (oxygen consumption) of neurons. Typically, following a
stimulus, one records a pattern in the signal known as the canonical hemodynamic
response function (HRF). To interpret the BOLD signal various templates of the
canonical HRF are generated from the experimental design of the stimuli and the
weights for each template are computed with a general linear model (GLM), these
weights are colloquially called “betas”. For the visual cortex, of particular
interest is the BOLD response (or betas) to naturalistic image stimuli. The
subject views a naturalistic image while in the scanner and then the signal is
recorded. This process can then be repeated for many images and the
resulting pairwise dataset can be used to train models that aim to estimate the
processing of visual information occurring in the brain. A prominent dataset is
the Natural Scenes Dataset (NSD) which will be used in this work
\cite{allen_massive_2022}.    

One may categorize CNN based fMRI models found in the literature as encoding
models, and decoding models. The former aims to predict the BOLD signal from an
image stimulus, and the latter models aim to achieve the inverse. Since
naturalistic images are quite complex the decoding task is more challenging and
has also been popular recently. Preceding CNNs, it was shown that simple
encoding models can be built with Gabor filters \cite{kay_identifying_2008}.
Remarkably, these models can also be used to select, from a novel image dataset,
the image a subject is viewing from just the corresponding fMRI signal. This is
done by selecting the image which when passed through the encoding model
correlates the most with the fMRI signal at hand \cite{kay_identifying_2008}.
This can be thought of as a rudimentary decoder. A relatively recent approach to
decoding that involves CNNs is taken by \cite{shen_deep_2019}: here a linear
model was used to predict features in all the layers of the pre-trained VGG19
model from the fMRI voxels, then a separate pre-trained Generative Adversarial
Network (GAN) was used to iteratively optimize the input to VGG19 until the
resulting feature vectors of the image matched the ones decoded from the fMRI
signal. Numerous similar approaches exist in the literature that use pretrained
CNNs, GANs, Variational Autoencoders (VAEs), etc., to perform decoding
\cite{beliy_voxels_2019,
du_reconstructing_2019,gaziv_self-supervised_2022,seeliger_generative_2018,st-yves_generative_2018}.
Reviews discussing architectural comparisons, benchmarks and outlooks are
\cite{rakhimberdina_natural_2021, du_fmri_2022}. Recently stable diffusion
models, which are known to provide better reconstructions than GANs
\cite{dhariwal_diffusion_2021}, have also been used to build decoding models
\cite{ozcelik_natural_2023, takagi_high-resolution_2023}. 

The focus of this study will be on CNN encoding models. The architecture we will
investigate is based on the work of \cite{st-yves_feature-weighted_2018} and is
straightforward; we use a pre-trained CNN, like AlexNet, as a feature extractor
or “backbone” and have a “readout” to the voxels. This readout is usually a
linear model from each pixel of the features to the voxels of interest. As noted
above we may think of the BOLD signal as a metabolic signature of the underlying
neuronal activity, since a typical voxel can contain millions of neurons,
different neuronal states can correspond to the same BOLD state of the voxel,
i.e., high degeneracy per voxel. The rationale is that the artificial neurons
(ANs) of the CNN serve as an estimate of the hidden neuronal state. A similar
approach is followed in \cite{guclu_deep_2015} where the authors created an
encoding model and showed that the hierarchical processing in a CNN was mapped
to the brain, revealing a gradient of complexity in downstream processing in the
ventral pathway. Some decoding approaches mentioned above also use an
encoder build in this manner
\cite{beliy_voxels_2019,gaziv_self-supervised_2022}. Similar approaches to
building encoders have also been used to demonstrate emergence of
non-hierarchical representations in the backbone CNN
\cite{st-yves_brain-optimized_2023}. Such models have also been used in
conjunction with text encoders to show how text captions of images help to
better predict high-level visual areas \cite{wang_better_2023}. 

We will also build models that take text captions of images as inputs and try to
predict the BOLD signal. For language processing the most prominent architecture
is that of transformers \cite{vaswani_attention_2017}, which are based on the
concept of attention. The relevant architecture is that of CLIP \cite{radford2021learning}, which takes in
a text string and an image as inputs. The text is mapped via a transformer to an
embedding space, while the image passes through a CNN or a vision transformer to
the same embedding space. This allows us to compare the similarity of text
descriptions and images. 

Our aim in this study is to put forth a rigorous interpretation of such models
by using XAI tools such as “dreaming” (i.e., input optimization)
\cite{olah_feature_2017}, and integrated gradients for pixel attribution
\cite{sundararajan_axiomatic_2017}. We
elucidate what effect the size of the feature space (number of CNN filters used)
has on the accuracy and mechanism of the prediction (pruning). Further, as noted
above, for a given BOLD state in a voxel there is significant degeneracy in the
underlying neuronal state, which is estimated by the artificial neurons (ANs).
So then, what should determine the state of these ANs? In this setup this is
determined by the downstream task, the architecture, and the pre-training dataset
of the backbone. As such, we build encoders with various backbones and use our
XAI toolbox to see how this affects the mechanism and accuracy of the encoder.
Finally, we may also use the pretrained state of the ANs as an initial condition
for the encoder and then optimize them during the training process to see how
this affects the mechanism and accuracy of the prediction, i.e., finetuning the
backbone.

\section{Methods}
\subsection{Data}
We use the Natural Scenes Dataset (NSD) to train our models
\cite{allen_massive_2022}. This dataset was created by recording the fMRI
responses of eight participants who viewed 9,000-10,000 distinct images of
natural scenes. The fMRI scans were performed using a 7T whole-brain
gradient-echo EPI at 1.8-mm resolution and 1.6-s repetition time. The images
were supplied from the Microsoft Common Objects in Context (COCO) database. We
use subjects that completed the full protocol which are subjects 1, 2, 5 and 7. For each subject, the models were trained
 using 8859 images and 24980 fMRI trials (up to 3 repetitions per image)
 and used 982 images and 2770 fMRI trials for testing. We
split the data into training and testing datasets in the default manner as
provided by the NSD code repository. Only voxels within the visual cortex were
used for training and testing. 
\subsection{Vision Encoding model}
Here we outline our encoding model which is related to the work of
\cite{st-yves_feature-weighted_2018}. We have two spaces to consider, the space
of voxels, $\mathcal{B}$, and the space of input images, $\mathcal{P}$. As
we go deeper in the layers of a CNN, some layers will decrease the resolution.
We have then for each layer of the network a pixel space, $\mathcal{P}_l
\subseteq \mathcal{P}_{l-1} …  \subseteq \mathcal{P}_{1} \subseteq \mathcal{P}_0$.
In general each of these spaces has a different pixel dimension due to the
changing resolution, and they will also have channel dimensions. Typical operations
that change the resolution are convolutions and pooling. Then, let
$|\phi^l_k[I]\rangle \in \mathcal{P}_l$ be a feature map for the $k^{\text{th}}$
feature in a layer, for an image, $I$. Here we are using “braket” notation for
vectors. Now we construct the readout to $\mathcal{B}$, for each voxel consider
some receptive fields (RFs) $\langle\rho^{\ell}_v|$, each of these RFs are
trainable weights with the same dimension as the pixel space of layer $\ell$,
denoted as, $\text{dim}_{\text{pixel}} (\mathcal{P}_\ell)$. These RFs per voxel
will not exist for every layer of the CNN, but rather at some predetermined
layers denoted with a different script, $\ell$. We have then,
\begin{equation}\label{eq:readout}
    \beta_v[I] = \sum_{\ell,k} w^{\ell}_{k,v} \langle \rho_v^\ell | \phi ^\ell_k [I]\rangle + b_v
\end{equation}
where, $w^\ell_{k,v}$ and $b_v$ are the trainable weights and biases
respectively, $\beta_v$ is the BOLD signal for the voxel $v$, and $\langle |
\rangle$ denotes an inner product over pixels. 

\begin{figure}[h]
    \centering
    \hspace{2.9cm}
    \includegraphics[scale=0.5]{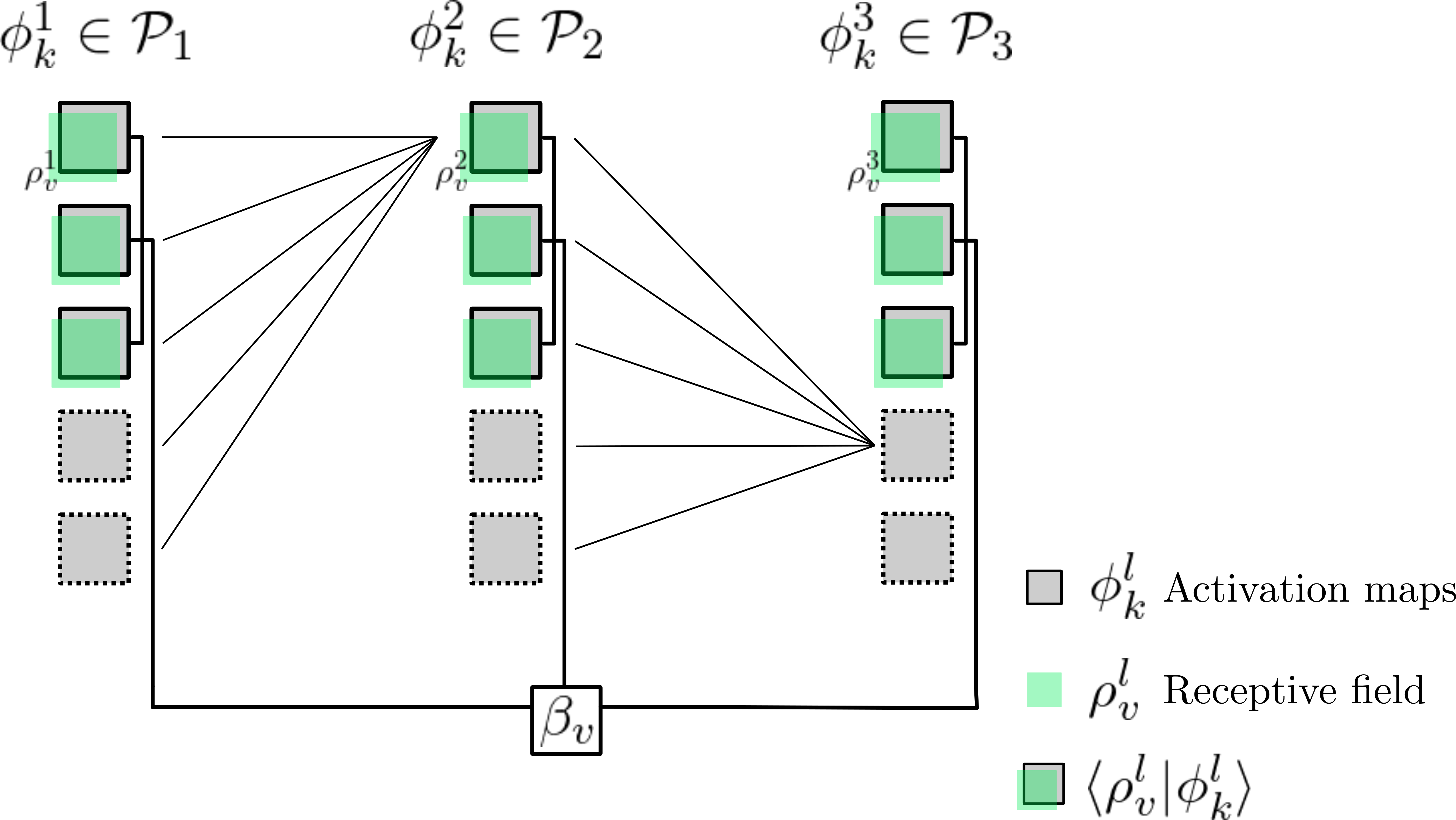}
    \caption{This is an illustration of \ref{eq:readout}. The grey and green squares, denote 
             activation maps, $\phi_k^l$, of each filter and the receptive field, $\rho_v^l$, for a voxel $v$, respectively.
             In this instance, the activation maps with dashed lines are not included in the readout, but they still contribute
             to activations for filters in the next layer. Not all connections between layers are shown for clarity. 
    }\label{fig:model}
\end{figure}

We can train this model in two
ways, we either only train the readout weights and don’t finetune the backbone
or we finetune the backbone. Three pre-trained CNNs are used, AlexNet
\cite{krizhevsky_imagenet_2012}, Vgg \cite{simonyan_very_2015}, and CLIP RN50x4 \cite{radford2021learning}
(referred to as RN50x4 hereafter). AlexNet is chosen for its simplicity, with
Vgg we move one step towards a larger feature space with deeper layers and
smaller filters. The visual branch of CLIP with RN50x4 architecture is chosen as
a model on the far end of the spectrum offering a large ResNet architecture
trained with millions of web scraped images. We use the Adam optimizer for
training with ten epochs at a learning rate of 0.0001. These parameters were
chosen by experimenting on subject 1 with various values. As mentioned above, we
have a choice in which layers and filters, $\phi^{\ell}_k$ are used in the sum
in \ref{eq:readout}. The layers chosen from each model along with the maximum
number of filters per layer are shown in \ref{tb:features}. For Alexnet and Vgg
we take readout from each of the ReLU layers. Since the
Resnet architecture is more complex we take readouts at the ReLU layers at the
end of each block. ReLU layers are a natural choice here due its simplicity in
returning a zero or  a positive activation. This gives us a response that
closely mimics an ON/OFF switch for each feature which can then be combined with
readout weights to create more complex combinations of features. The number of
filters per layer is a parameter that is varied; we compute the activations of
all the images in the training set and sort, in descending order, the filters in
each layer based on their standard deviation. Then we choose the top $p$\% of
the total filters for the readout. Accuracy is quantified per voxel as a
correlation with the ground truth signal, where this correlation is computed
over all test images.

\begin{table}[h]
    \centering
    \begin{tabular}{|l|c|l|c|l|c|}
        \hline
        \multicolumn{2}{|c|}{Alexnet} & \multicolumn{2}{c|}{Vgg11} &
        \multicolumn{2}{c|}{RN50x4} \\
        \hline
        Layer name & \# filters & Layer name & \# filters & Layer name & \#
        filters \\
        \hline
        features.2      & 64    & features.2    & 64    & layer1.3.relu3  & 320
        \\ 
        features.5      & 192   & features.5    & 128   & layer2.5.relu3  & 320
        \\ 
        features.7      & 384   & features.7    & 256   & layer3.9.relu3  &
        1280\\ 
        features.9      & 256   & features.9    & 256   & layer4.5.relu3  &
        2560\\ 
        features.12     & 256   & features.12   & 512   &                 & \\ 
                        &       & features.15   & 512   &                 & \\ 
                        &       & features.17   & 512   &                 & \\ 
                        &       & features.20   & 512   &                 & \\ 
        \hline
        \textbf{Total}  & 1152  &               & 2752  &                 &
        4480\\ 
        \hline
    \end{tabular}
    \caption{This table summarizes the layers choosen for the readout and the total number of filters in each layer.}\label{tb:features}
\end{table}

\subsection{Text encoding model and word clouds}
Each image in the NSD dataset also has five captions associated with it
\cite{allen_massive_2022}. These captions along with a pre-trained transformer
model can be used to predict the bold signal \cite{wang_disentangled_2023}. The
text encoding model we use here is straightforward, we take the last layer of
the CLIP text encoder and add one layer of weights to map it to the voxels. We
do not attempt to finetune the transformer.

We also make use of word clouds to illustrate to which words the images are most
correlated with. This is done by choosing an ensemble of common words, we use
the Brown Corpus \cite{francis1979brown} to create an ensemble of top 10,000
common words (nouns) via the NLTK library \cite{bird2009natural}. These are then
passed through the CLIP text encoder to get the common text/vison embedding
vector. From here we can pass images to the vision encoder of CLIP and compute a
similarity score with the embedding vector of each of the words. Top 20 words
are shown as word clouds where the font size is proportional to the similarity
score. 

\subsection{Attribution}
The basic idea behind attribution is to determine what drives the activation of
a particular AN or voxel. We consider pixel attribution: a measure of the
location of the features in the pixel space that drive the activation. To
compute pixel attribution, we use the integrated gradients approach
\cite{sundararajan_axiomatic_2017} which can be summarized as follows. Let us
denote the activation of a given AN or voxel of interest as $F$, then for
integrated gradients we define the following quantity,
\begin{equation}
     G(F)_i = \int_0^1 \frac{\partial}{\partial I_i} F \left(I’ + \alpha (I - I’) \right) d\alpha 
\end{equation}
Here, $I_i$ is the $i^\text{th}$ pixel of the input image and $I’$ is a baseline
image which we take to be zero. The integration here is to take into account the
fact the gradient will change based on the magnitude of the argument. We average
the color dimension $G(F)_i$ and apply a smoothing filter with sigma of two pixels to
reduce noise. 

\subsection{Maximally exciting images (MEIs)}

Maximally exciting images (MEIs) or “dreams” are images that most excite an AN
or voxel. Dreams, in this context, are synthetic images created by input
optimization. One starts with random noise as an input image and then
iteratively updates this image towards one that invokes high activation in an
AN.   
We can also generate MEIs for arbitary loss functions; the loss function we
utilize the most is simple where the goal is to maximize the mean beta value for
an ROI,
\begin{equation}
    I_0=\mathop{\arg \max}\limits_{I} \bar{\beta}_{\text{ROI}}[I]
\end{equation}
where, $I_0$ is the MEI and $\bar{\beta}_\text{ROI}[I]$ is a mean over the
voxels of a particular ROI. The MEIs are generated from a noise starting
condition and MSE denotes the mean squared error, details can be found in
\cite{kiat_greentfrapplucent_2024, olah_feature_2017}. 

We adapted the lucent library \cite{kiat_greentfrapplucent_2024} for CNNs, so
that it can be used to generate MEIs for voxels. In
\cite{walker_inception_2019}, authors have used ‘inception loops’ to show how
dream MEIs invoke activation in target cells of mice that is significantly
higher than dataset MEIs. Dream MEIs tend to be abstract; other approaches, like
GANs, have been used to generate dream MEIs that are more naturalistic
\cite{gu_neurogen_2022}. Further, it has been shown that such MEIs modulate
brain responses \cite{gu_human_2023}.

\section{Results}
\subsection{Accuracy}
\begin{figure}
    \centering
    \includegraphics[width=0.77\linewidth]{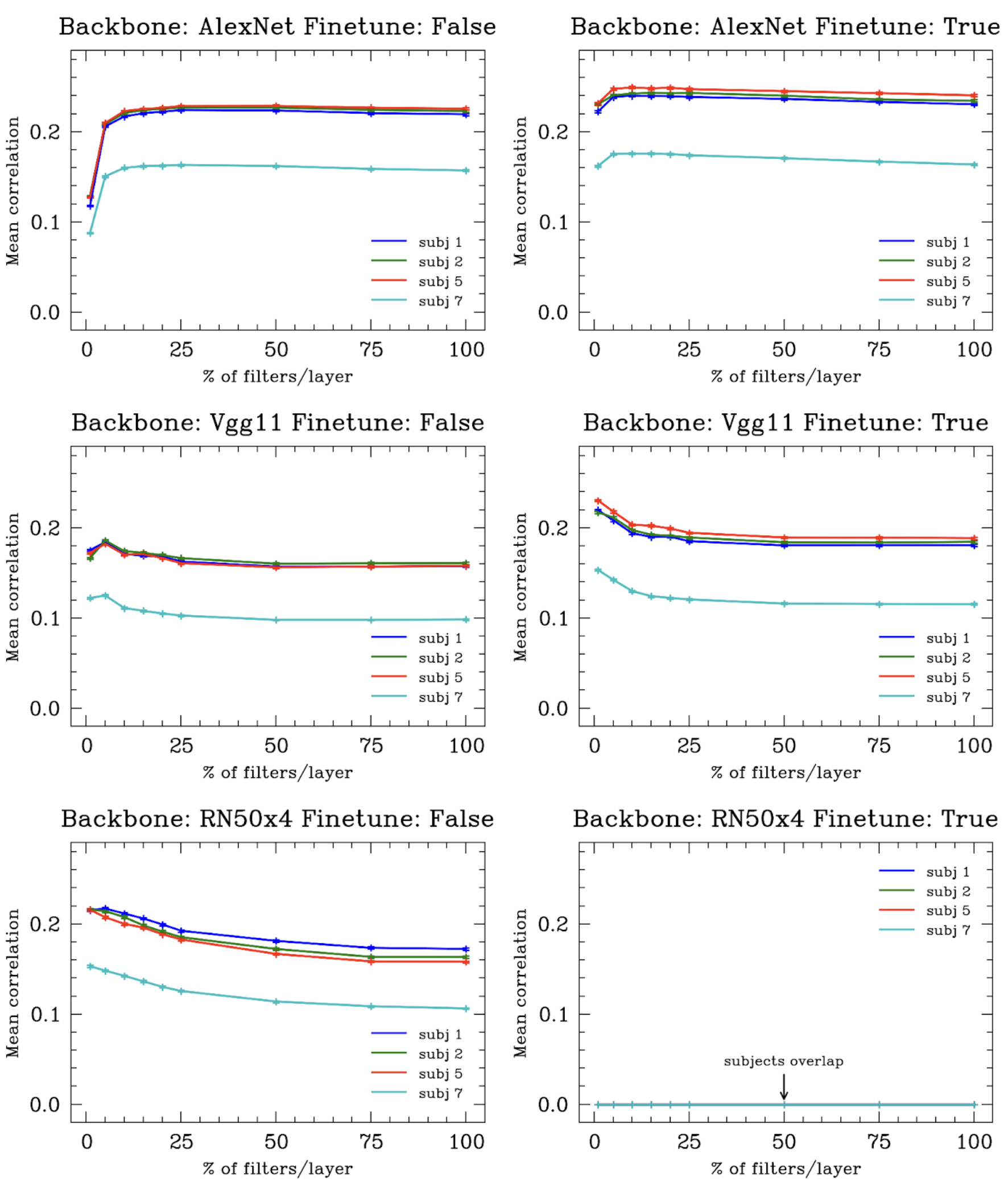}
    \caption{This figure shows the trend in the correlation with the ground
            truth as a function of the percentage of filters per layer. The
            y-axis is the mean correlation over all voxels, the x-axis is the
            percentage of filters per layer parameter and the points indicate
            the models we trained. The plots also contain error bars which show
            the standard error. Different subjects are colour coded as indicated
            in the legend. Each row is for a particular backbone and the columns
            are for whether the backbone was finetuned or not. }
    \label{fig:accuracy}
\end{figure}

The parameter used to prune is the percentage of filters per layer chosen to
read out to the voxels. \ref{fig:accuracy} shows the results of changing this parameter. The
rows show each backbone while the columns show the plots for whether or not the
backbone was fine tuned. The y-axis is the mean of the correlation taken over
voxels. We sample the percentage parameter as (1, 5, 10, 15, 20, 25, 50, 75,
100), this uneven sampling is done in order to have more points at low values
where we see some changes. There are error bars accompanying the points which
show the standard error. Overall it is clear that the size of feature space per
layer has very little effect on the correlation. Also, subject 7 appears to be
an outlier. Attempting to fine tune the RN50x4 backbone with the same scheme as
the other backbones results in zero or NaN correlations and has been left out in
following plots. Comparing all the cases, the fine tuned Alexnet backbone with
10 percent of filters per layer performs the best, although only marginally. 

\begin{figure}
    \centering
    \includegraphics[width=0.999\linewidth]{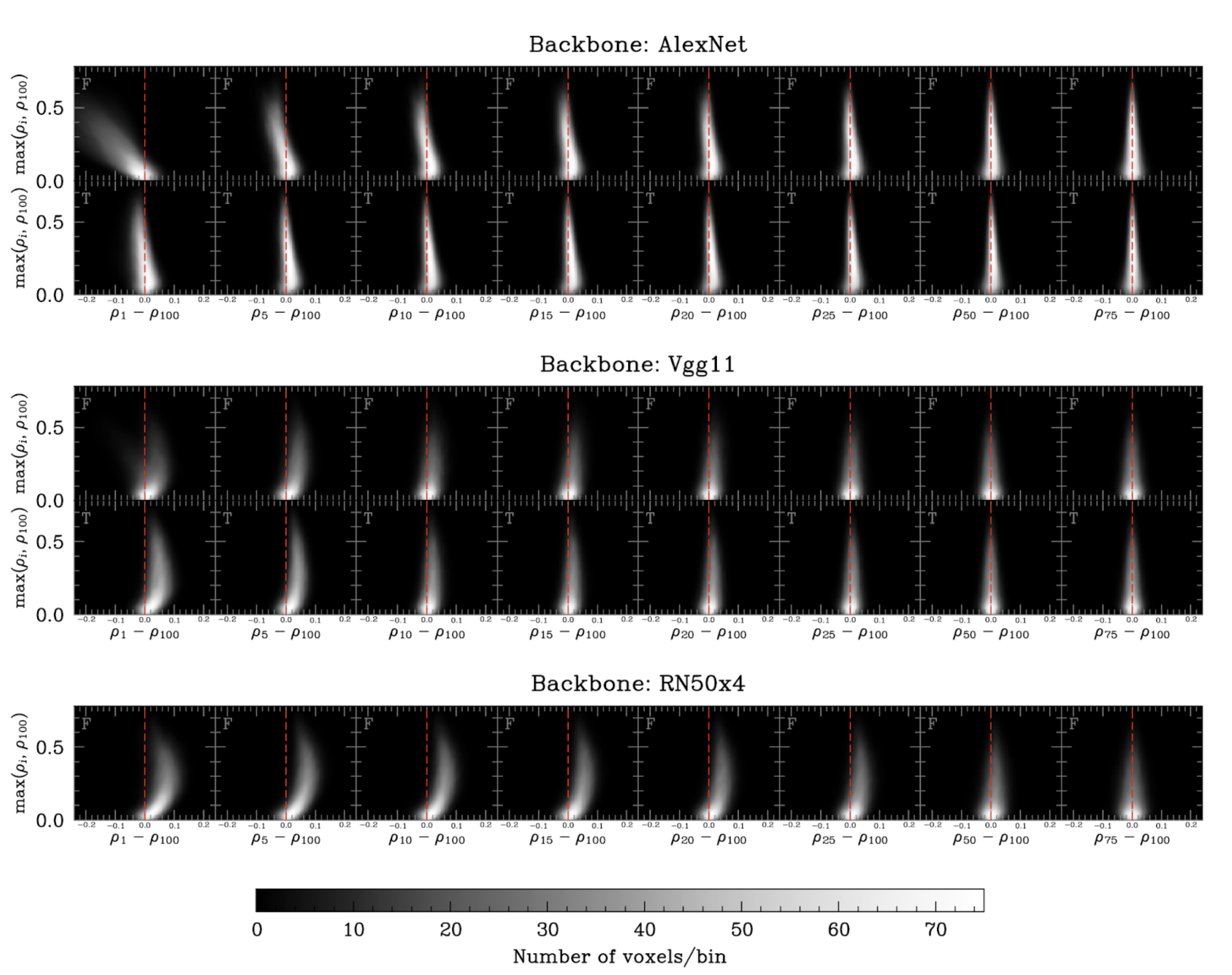}
    \caption{This figure aims to demonstrate the voxel distribution as a
            function of the percent parameter for each of the backbones. The two
            rows within each panel shows whether or not the model was fine
            tuned, this is denoted with a “F” (no-finetuning) or a “T”
            (finetuning) on the top left corner of each plot. In each small
            panel the x-axis shows the difference between $\rho_i$ and
            $\rho_{100}$, where $\rho_i$ is the correlation using $i\%$ of the
            filters. The y-axis is the maximum correlation out of $\rho_i$ and
            $\rho_{100}$. The intensity at each point denotes the number of
            voxels that fall into that bin as shown with the bar at the bottom.
            If the intensity in these plots moves to the left of the red dashed
            line, the model with 100 percent of the filters performs better.}
    \label{fig:flames}
\end{figure}

\ref{fig:flames} shows the distribution of the correlation values of the voxels for each
pruned model, while also performing a comparison with the unpruned version. The
rows here are for each backbone, each row is split into two where the top one
contains results where the model was not finetuned and the bottom contains
results for when the model was fine tuned. This is indicated on the top-left of
each plot with “F” or “T”, respectively. In each small plot in Fig 2 the x-axis
contains bins for the difference, $\rho_i - \rho_{100}$, where $\rho_i$ is the
correlation calculated using $i\%$ of the filters per layer. The y-axis is bins
for the maximum correlation between the two. The red vertical dashed line marks
$\rho_i = \rho_{100}$, if more intensity falls to the left of this line then the
model with 100\% of the filters performs better. The bar at the bottom shows the
relationship between the intensity and the number of voxels. Voxels are taken
from all subjects except subject 7. This provides us with a birdseye view of the
averages taken in \ref{fig:accuracy}. We can see that the AlexNet backbone has brightness
distributed relatively more evenly along the “flame” as compared to the other
backbones where there is a concentration at the bottom. The AlexNet intensity
also reaches slightly higher indicating a higher max correlation. The trends
shown in \ref{fig:accuracy} are manifest in these plots, for example, for AlexNet with no
finetuning we start with the intensity to the left of the redline moving towards
a more symmetric distribution as the percent parameter increases. For AlexNet
with finetuning we see a subtle shift of the intensity to the right which peaks
at 10\% and then moves towards a more symmetric distribution.          

\begin{figure}
    \centering
    \includegraphics[width=0.999\linewidth]{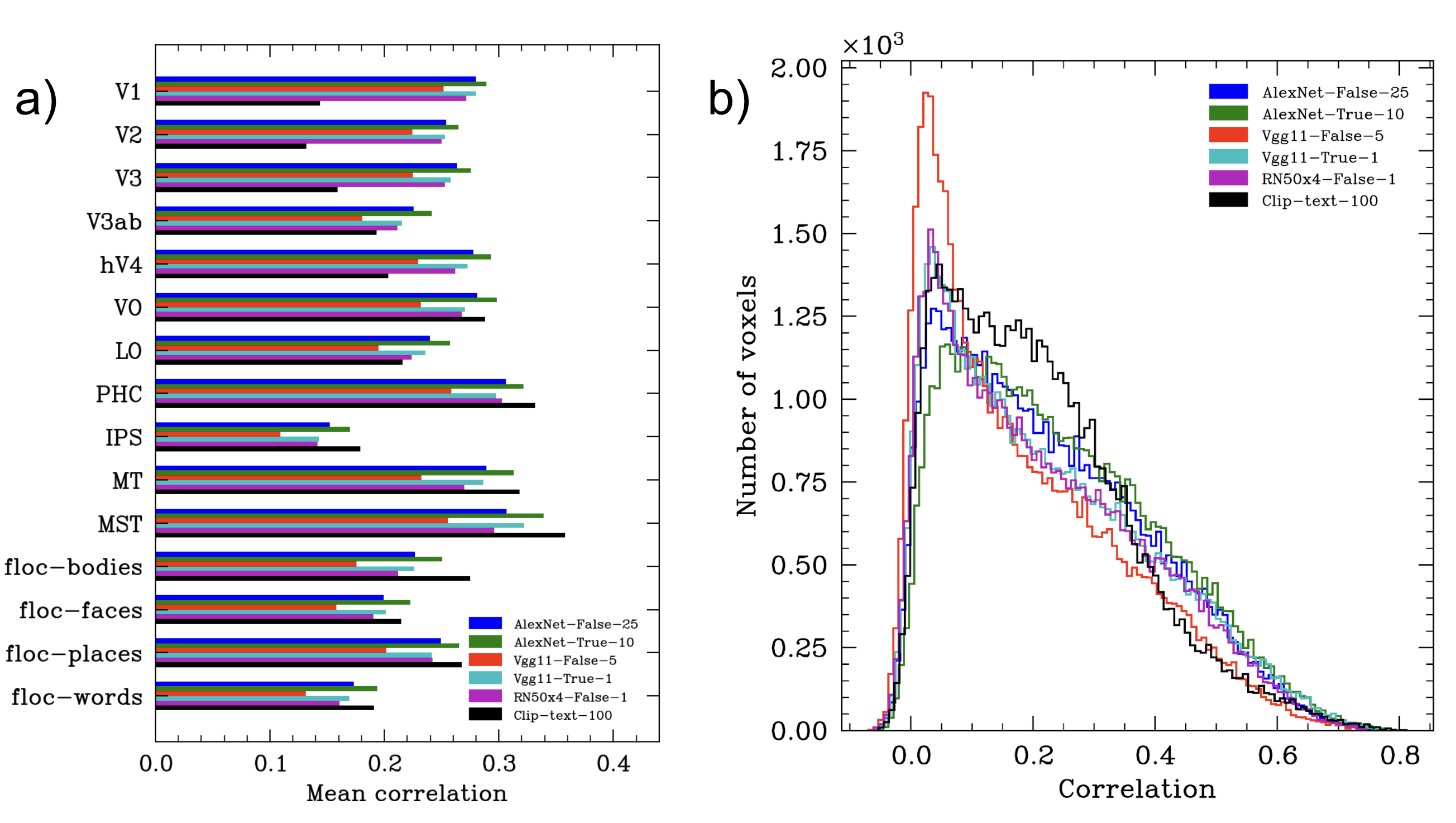}
    \caption{In these plots we leave out subject 7 since it is an outlier. Panel
             a) shows the mean correlation over voxels in each ROI, for each
             model that performs well in its class. The text encoder is also
             included here (shown with black). The legend shows the
             corresponding model for each colour, arranged as
             <backbone>-<finetune>-<percent>. In panel b) the x-axis represents
             correlation value bins and y-axis shows the number of voxels that
             fall into that bin, again the colours show different models.}
    \label{fig:hist}
\end{figure}

In \ref{fig:hist}a) we have selected the best percentage parameter for each model and
calculated the mean of the correlation over voxels from all subjects except
subject 7 in a particular ROI, as illustrated with different colors which are
defined in the legend. Overall we see that the finetuned AlexNet backbone with
10\% of filters per layer (green) performs the best. As noted by
\cite{wang_better_2023} the text encoder (black) performs slightly better in
higher visual areas, and significantly lower in early visual cortex areas, but
the increase is marginal.  

In  \ref{fig:hist}b) the x-axis shows correlation bins and the y-axis is the number of
voxels that have a correlation that falls into that bin. All subjects except
subject 7 are combined to make this histogram, the color coding is the same as
panel a). Again, we see that the finetuned AlexNet backbone with 10\% of
filters per layer (green) performs the best but other models are similar.

\subsection{Dreams}
\begin{figure}
    \centering
    \includegraphics[width=\linewidth]{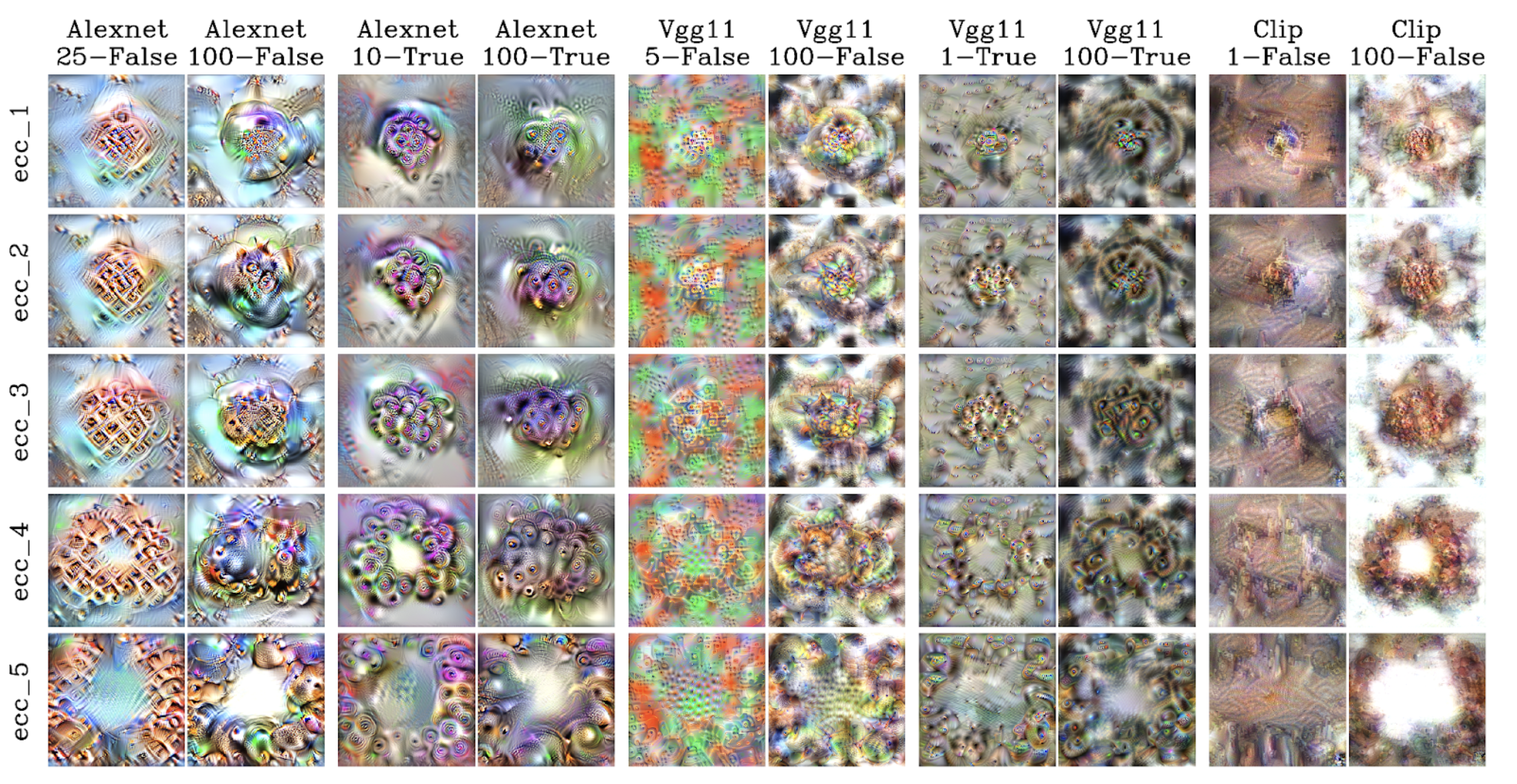}
    \caption{This figure shows the results for subject 1 of the dreams for the
    eccentricity ROIs (row) for the best backbones and the corresponding one
    with 100\% filters (columns). Each column is a backbone specification
    denoted in the title as <backbone>-<percent>-<fine tuning>.   
    }
    \label{fig:ecc_subj-1}
\end{figure}

In \ref{fig:ecc_subj-1} we see the result of the dreams (MEIs) for the retinotopy eccentricity
ROIs for subject 1. The columns show the best backbones and the
corresponding one with 100\% filters. The rows are the different eccentricity
ROIs moving radially outward. We see that although the features are different
there is a general overall trend of a ring of features moving outward as we go
from ecc\_1 to ecc\_5. Interestingly a lot of these features appear to be
circular. 

\begin{figure}
    \centering
    \includegraphics[width=\linewidth]{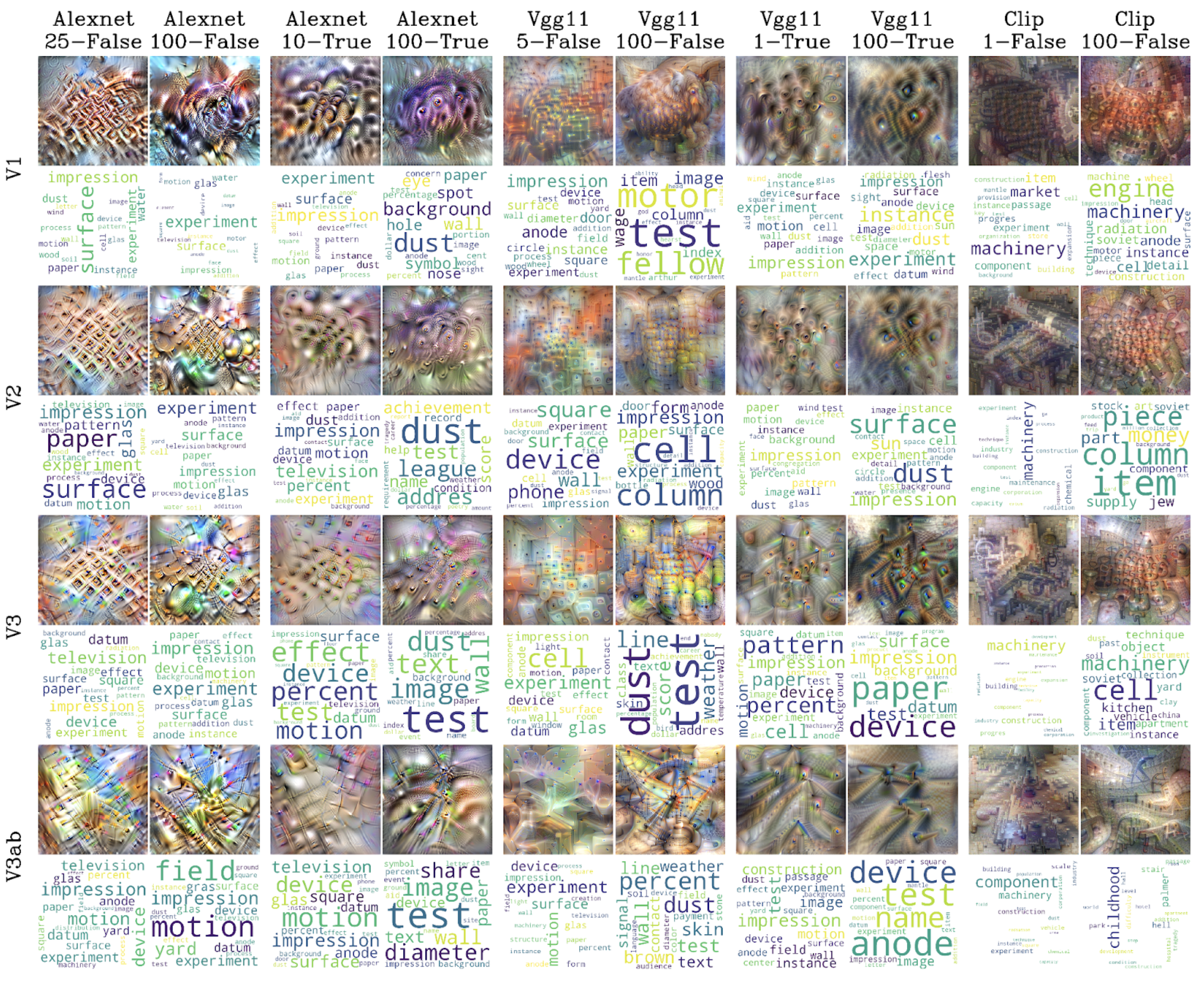}
    \caption{This figure shows the results for subject 1 of the dreams, with word clouds underneath, for the early visual cortex areas V1,V2, V3 and V3ab (rows) for the best backbones (columns). The size of the words denote the similarity with the image.  }
    \label{fig:early_subj-1}
\end{figure}

In \ref{fig:early_subj-1} we generate dreams and word clouds for some early areas of the visual
cortex for subject 1. The columns are the same as \ref{fig:ecc_subj-1}. The rows correspond to
these areas, V1, V2, V3 and V3ab, each row is split into two, with the lower one
showing the word clouds. The size of a word in the cloud illustrates the
similarity of the word with the image. Generally we don’t see any particular
pattern to the words. The dreams are quite dependent on the backbone used.

\begin{figure}
    \centering
    \includegraphics[width=\linewidth]{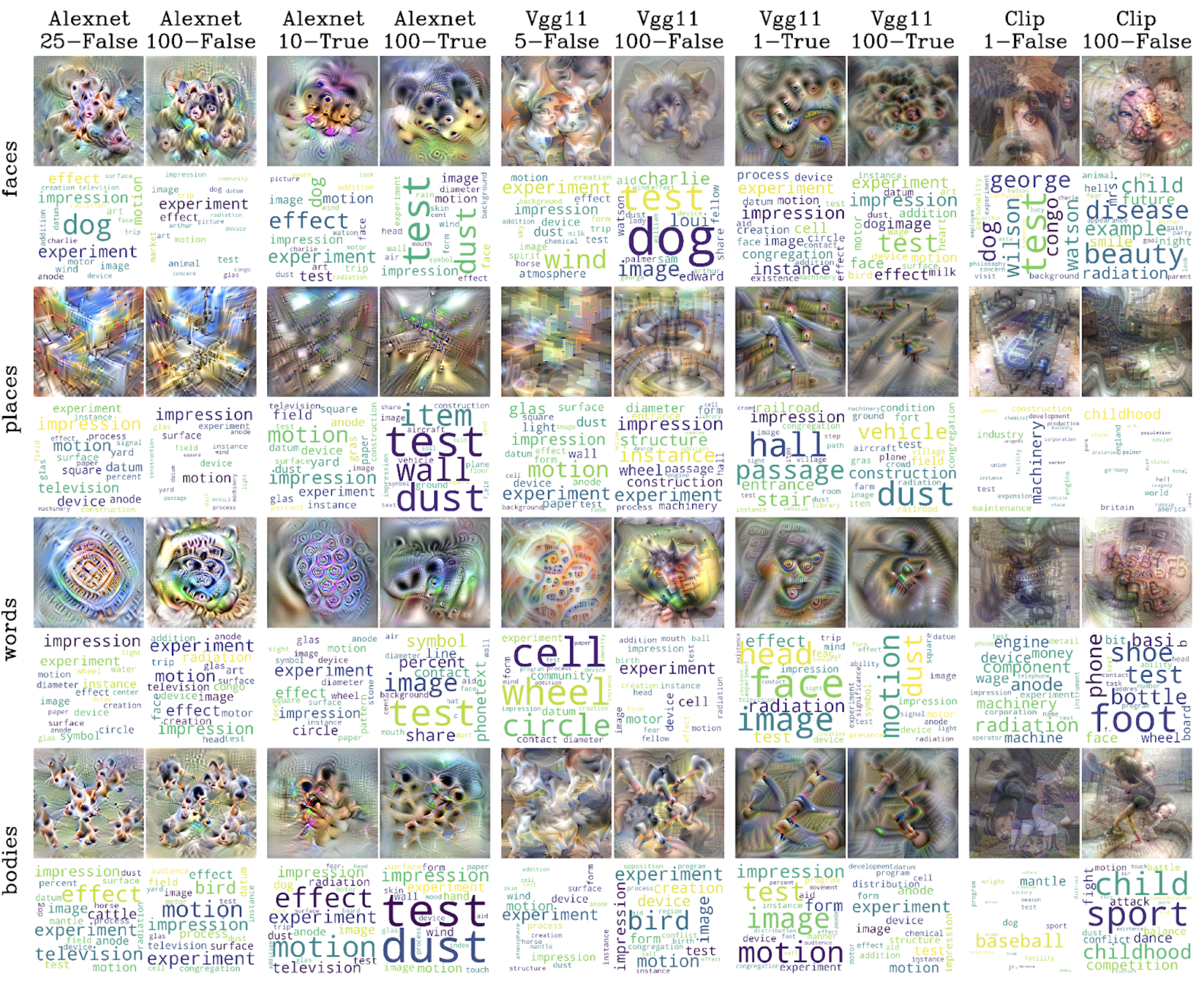}
    \caption{This figure shows the results for subject 1 of the dreams, with word clouds underneath, for the higher visual cortex areas corresponding to faces (OFA, FFA, mTL-faces and aTL-faces), places (OPA, PPA and RSC), words (OWFA, VMFA, mfs-words and mTL-words), and bodies (EBA, FBA and mTL-bodies) for the best backbones (columns). The size of the words denote the similarity with the image.}
    \label{fig:late_subj-1}
\end{figure}

In \ref{fig:late_subj-1} we generate dreams and word clouds for higher visual areas of the
visual cortex for subject 1. The rows show these ROIs, faces contains the areas
OFA, FFA, mTL-faces and aTL-faces, places contains the areas OPA, PPA and RSC,
words contains the areas OWFA, VMFA, mfs-words, and mTL-words, and bodies contains the areas EBA, FBA and
mTL-bodies. Here in the word cloud we can see some relevant words, for the faces
row we see the appearance of relevant words like “dog, face, animal, child,
head, smile”. Although abstract one can subjectively discern features that
relate to faces, this is especially true for the not finetuned AlexNet and Vgg11
backbones. When we finetune these backbones we see features that closely
resemble eyes. The CLIP backbone produces animal-like features when using 1\% of
the filters per layer. Although we clearly see human face features when using
100\% of the filters. In the second row we have the dreams for the places ROI.
Here we see the appearance of some relevant words like “yard, wall, structure,
hall, passage, construction, England, Germany, America, China, hell”; a lot of
the names of countries appear when using the CLIP backbone. Subjectively, the
visual features for AlexNet and Vgg11, resemble structures, buildings, passages.
The CLIP 1\% dream seems to resemble a factory floor with machinery (also
included in the word cloud). The CLIP 100\% dream is quite elaborate and
relevant, where we can see features that resemble architecture, passages and
grass. Moving down we have the words ROI, here the words in the word cloud are
random. The visual features for AlexNet and Vgg11 show similarity to early areas
shown in \ref{fig:early_subj-1}. Remarkably, for the CLIP backbone with 100\% of filters we can
actually see some letters. Finally, in the last row we have the bodies ROI which
is made from the areas EBA, FBA and mTL-bodies. The word clouds for AlexNet and
Vgg11 show random words except perhaps for motion, bird, congregation and form. The
CLIP word clouds show baseball, sport, competition, these are more relevant
coming from images of people playing sports. The visual features are quite
abstract, although for the CLIP backbone with 100\% of the filters we see
features that resemble limbs.

\subsection{Implicit Attention}
\begin{figure}
    \centering
    \includegraphics[width=\linewidth]{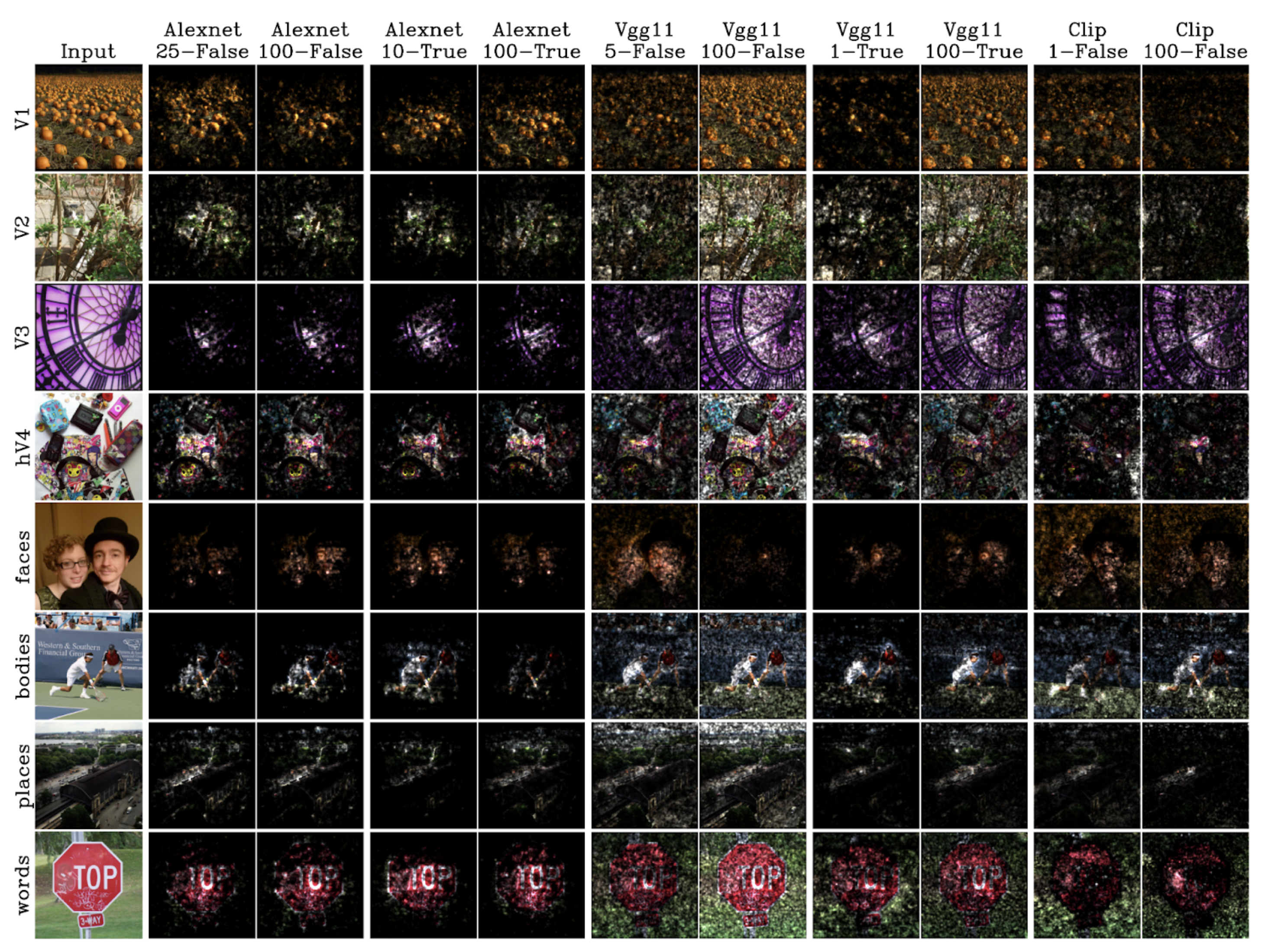}
    \caption{This figure shows the implicit attention from the integrated gradient approach as an intensity mask. The first column is an input image, and the remaining columns are the different backbone configurations, titled as <backbone>-<percent>-<fine tuning>. The rows show the different ROIs.   }
    \label{fig:attn_subj-1}
\end{figure}

In \ref{fig:attn_subj-1} we see the results of the maps generated by the integrated gradient
approach (other subjects in the \ref{Appendix}). The rows show the different ROIs and the columns are the different
backbone specifications. The maps are illustrated as intensity masks. This
approach gives us insight into which image features are contributing the most
towards the signal in a voxel. We can see that there is quite a bit of
variability here indicating different mechanisms of prediction dependent on the
backbone chosen. For example, we can see that Vgg11 and CLIP tend to focus on
the background also and not just the central region as is the case for AlexNet.

\section{Discussion}
In this work we have comprehensively dissected CNN based fMRI encoding models of
the visual cortex. Perhaps the most central quality of CNNs is their
hierarchical processing of features; low-level features are combined to produce
higher level features. As an illustration of this concept consider a rudimentary
detector of square shapes. In the first layer, filters with a small receptive
field will look for horizontal and vertical edges. Following a pooling step, the
next layer might introduce a filter that gets activated by a specific
arrangement of these activations from the first layer: two horizontal and two
vertical edges precisely spaced to form a square. Naturally, this process gets
extremely complex with depth, but this toy example helps us understand intuitively how
hierarchical processing combines features as the effective receptive field
grows.

Consequently, this means that filters interact closely, and these
interactions are, by design, governed by the downstream task and the training
dataset. For AlexNet and Vgg11 this is a classification task on the ImageNet
dataset, and for the CLIP RN50x4 model the goal is alignment with captions with
image-caption pairs curated from the web. The general idea of the fMRI encoding
model is to intercept the feature evolution at various levels of the
hierarchical processing and then map it to voxel space with \ref{eq:readout}. As shown in
the literature in many sources (for a review see
\cite{rakhimberdina_natural_2021}) and corroborated by our accuracy plots, this
approach demonstrably encodes the fMRI signal in a non-trivial manner. However,
given the immense complexity inherent in CNN architectures the precise
mechanisms of prediction are challenging to fully decipher. In this work we have 
addressed some of these challenges. 

The eccentricity ROI dreams serves as a litmus test to examine if the encoding
model can generate stimuli that are reasonably close to retinotopy experiment
stimuli \cite{engel_retinotopic_1997}. \ref{fig:ecc_subj-1} shows the results for subject 1
(other subjects in \ref{Appendix}), we can see clearly that with increasing eccentricity an
abstract ensemble of natural features that start at the center and move radially
outward. Typically, these are done with concentric rings filled with a
checkerboard pattern. Here, we observe distinct features for each backbone.
However, within a given backbone, similar features emerge across eccentricity
ROIs. Additionally, different subjects exhibit similar features for the same
backbone. This pattern suggests that the signal effectively encodes the
"scaffolding" of the features (i.e., the rings), while the finer details and
overall style—reminiscent of the concept of style in style transfer
\cite{jing2019neural}—are not strongly constrained by the signal. As we will
discuss below, a similar mechanism operates in higher-level brain areas as
well.

With BOLDreams we can construct encoding models with various backbones as
demonstrated in this study. We tried three different backbones, we toggled finetuning,  
and adjusted the width of the readout (percent of filters per layer).
In all these combinations we find that the changes in correlation are marginal.
This is clearly seen with the histogram in \ref{fig:hist}b), in the right tail the
visual backbones perform similarly. The shape of the text model (CLIP
transformer) is different as it performs poorly on the early visual cortex
\cite{wang_better_2023}. Clearly, adding more features does not improve the 
accuracy of the model. Interestingly, the finetuned version of AlexNet with 
just 1\% of filters per layer also performs reasonably well (\ref{fig:accuracy} 
and \ref{fig:flames}) and with 10\% of finetuned filters per layer we get the best 
model. This low number of filters provides insight into the complexity of
the feature space needed to encode the signal.

When we look at the dreams and implicit attention maps of different areas in the
brain we see that these models predict the signal in different ways. Each
backbone, and its configuration, has a different mechanism of prediction to
arrive at the BOLD signal. This is easily seen if we take extreme limits of this
spectrum, say comparing the dreams in the faces ROI of a complex model like the
CLIP RN50x4 backbone to AlexNet in \ref{fig:late_subj-1} for subject 1 and other subjects found
in the \ref{Appendix}. For Alexnet we see features that closely resemble animal features for
models that are not finetuned, we see “dog” show up often in the word clouds.
Whereas looking at CLIP we see very complex features; although somewhat
subjective, human-like features can be seen; faces, eyes, and skin. This
contrast in feature complexity between backbones is more apparent for the places
ROI. Here, for AlexNet and Vgg we see very rudimentary features like edges,
corners, paths — features one would expect are associated with cityscapes.
However, for the CLIP backbone we are able to see quite a bit of detail, we can
see passages, mountains, buildings, machinery, trees, etc. The word cloud also
gives plausible results like country names, valley, passage, apartment, village,
area, city, etc. This contrast in features continues for the words ROI, we see
that CLIP generates letters in its dreams, whereas AlexNet and Vgg11 show
rudimentary edge detectors. The implicit attention maps in \ref{fig:attn_subj-1}, also show
differing maps based on backbones. As with the retinotopic maps discussed above,
we observe that the signal captures a general concept, but the detailed features
within that concept are not strongly constrained. Different backbones tend to
default to features originally learned from their respective downstream tasks.
These features align with the concept dictated by the statistics of naturalistic
images originally found in the training set of each backbone, hence, the CLIP model
contains more detailed features.

Although the dreams of CLIP are more elaborate, and much simpler for AlexNet,
the prediction metrics for the two are hardly different. This suggests that adding more
features in the backbone does not improve the prediction of the BOLD signal. Conversely,
 adding more filters also doesn’t make the predictions worse.
This raises an intriguing question: What visual information does the BOLD signal
actually contain? This is particularly perplexing, as even AlexNet, with
fine-tuning and only 1\% of features per layer, successfully encodes the signal (\ref{fig:flames}).

One line of reasoning that fits this kind of model behavior is the following:
For the low-level brain areas simpler filters suffice to encode the BOLD signal. The
higher level brain areas have a behavior that resembles category classifiers, the
features that are sufficient to distinguish the classes are plenty to
encode the signal. However, in a large pre-trained model like the visual branch
of CLIP these basic broad features — a form of scaffolding — are connected to a variety of filters that provide finer details.
However, the presence or absence of these details does not impact the
predicted signal.

For AlexNet we observe the presence of animal features because classifying animals
is a downstream task. Whereas in CLIP the training data is much more diverse and
perhaps represents a better sample from the distribution of natural images,
offering many more connections providing richer details to the broad
features. 
We may also see this from a language perspective, the artificial
neuron that responds to the general concept of ``face'' will also respond to ``a
happy face'' or more complex statements like, ``God has given you one face, and you make yourself another'', the
readout will only assign high weights to the level of abstraction caught by the
BOLD signal which could be just ``face''. 

In the models we analyzed, this level of abstraction is satisfied by pruned readouts in AlexNet.
 This still doesn’t rigorously answer the question of exactly
what visual information does the BOLD signal have? However, our results indicate
that more rigorous and thoughtful analysis is needed in this regard, especially
when dealing with large pre-trained models that are somewhat of a mystery
themselves. Crucially, our findings point towards the
existence of a ``maximally minimal'' model, $M_0$: the model with the
least parameters that sufficiently encodes the BOLD signal. 

One subtlety to note about $M_0$, is that at a particular layer $l$ we may have filters ${\phi}^l_{k'}$
that are not contributing to the readout due to a narrower width. But they still contribute to a filter, $\phi_{k}^{l+1}$, in the next 
layer, perhaps strongly. This next filter could be part of the readout, these filters are illutstrated with a dashed border in \ref{fig:model}. 
This means that the signal is not strongly capturing certain low level features which 
are neccessary to build and encode higher level features which the signal does capture. 
This then is an example of how the signal in high level brain areas provides
some evidence for possible hidden neuronal states in lower level areas. Thus, $M_0$ would provide us with the foundational model that encodes the signal.
More evidence for hidden states can arise from other ``compatible'' connections,  $M_1$, that do not drastically 
affect the accuracy of the predicted BOLD signal. In such a decomposition, $M = M_0 \oplus M_1$, where $\oplus$ is a systematic fusion
of the two, $M_1$ would need external evidence to form connections, such as
evidence from other data modalities like, electrophysiology or behavioral
data, prior knowledge about neural architecture, or in the case of pretrained models a downstram task such as 
classification, text-image alignment (CLIP) or perhaps image reconstruction. For the pretrained models
extra connections are provided based on the statistics of natural images, i.e., conditional probabilities
of image features as derived from human generated captions. For example, if we consider faces, very simple
features can establish the existence of a face. However caption embedding would provide many more details regarding faces
that would point towards hidden neuronal states which are not constrained by the BOLD signal in NSD, as demonstrated by these results.

Incorporating XAI techniques into fMRI research workflows can help us design interesting experiments. Large pretrained 
models like CLIP can identify finer features that are compatible with higher visual areas, in the sense described above.
Dreams or images that strongly excite filters for these finer details can then be used to conduct further fMRI experiments in an attempt
to identify the region in the brain where neurons corresponding to those features would exist. Although ambitious, with 
robust XAI tooling, like BOLDreams, such dreams can be potentially generated in real time during the experiment. 

\section{Conclusion}
In this work we provide an open source XAI toolkit, BOLDreams, for training and
interpreting fMRI encoding models based on pre-trained text and vision
backbones. We have comprehensively dissected models with different backbones,
performed fine tuning, and pruning of readouts of such models. Our analysis
reveals that there is significant heterogeneity in the mechanism of prediction
of the fMRI signal which depends on the training data, architecture and the
downstream task of the backbone. We argued that this arises due to the
degeneracy of artificial neuron states that correspond to a fixed fMRI state,
which is evidenced by different specifications of the backbones showing similar
accuracies. Our analysis places emphasis on the search of a specialized
framework to identify the maximally minimal model that sufficiently encodes the
BOLD signal. Such a model can help design experiments that can help identify feature specific brain areas.

\section{Code availability}
The open-source repository for BOLDreams is available at \href{https://github.com/uhussai7/boldreams}{https://github.com/uhussai7/boldreams} 
where further details and documentation can be found.

\printbibliography

@article{allen_massive_2022,
	title = {A massive {7T} {fMRI} dataset to bridge cognitive neuroscience and artificial intelligence},
	volume = {25},
	copyright = {2021 The Author(s), under exclusive licence to Springer Nature America, Inc.},
	issn = {1546-1726},
	url = {https://www.nature.com/articles/s41593-021-00962-x},
	doi = {10.1038/s41593-021-00962-x},
	language = {en},
	number = {1},
	urldate = {2022-12-20},
	journal = {Nature Neuroscience},
	author = {Allen, Emily J. and St-Yves, Ghislain and Wu, Yihan and Breedlove, Jesse L. and Prince, Jacob S. and Dowdle, Logan T. and Nau, Matthias and Caron, Brad and Pestilli, Franco and Charest, Ian and Hutchinson, J. Benjamin and Naselaris, Thomas and Kay, Kendrick},
	month = jan,
	year = {2022},
	note = {Number: 1
Publisher: Nature Publishing Group},
	pages = {116--126},
}

@article{st-yves_feature-weighted_2018,
	series = {New advances in encoding and decoding of brain signals},
	title = {The feature-weighted receptive field: an interpretable encoding model for complex feature spaces},
	volume = {180},
	issn = {1053-8119},
	shorttitle = {The feature-weighted receptive field},
	url = {https://www.sciencedirect.com/science/article/pii/S1053811917305086},
	doi = {10.1016/j.neuroimage.2017.06.035},
	language = {en},
	urldate = {2022-12-20},
	journal = {NeuroImage},
	author = {St-Yves, Ghislain and Naselaris, Thomas},
	month = oct,
	year = {2018},
	pages = {188--202},
}

@article{gaziv_self-supervised_2022,
	title = {Self-supervised {Natural} {Image} {Reconstruction} and {Large}-scale {Semantic} {Classification} from {Brain} {Activity}},
	volume = {254},
	issn = {1053-8119},
	url = {https://www.sciencedirect.com/science/article/pii/S105381192200249X},
	doi = {10.1016/j.neuroimage.2022.119121},
	language = {en},
	urldate = {2022-12-20},
	journal = {NeuroImage},
	author = {Gaziv, Guy and Beliy, Roman and Granot, Niv and Hoogi, Assaf and Strappini, Francesca and Golan, Tal and Irani, Michal},
	month = jul,
	year = {2022},
	pages = {119121},
}

@inproceedings{beliy_voxels_2019,
	title = {From voxels to pixels and back: {Self}-supervision in natural-image reconstruction from {fMRI}},
	volume = {32},
	shorttitle = {From voxels to pixels and back},
	url = {https://proceedings.neurips.cc/paper/2019/hash/7d2be41b1bde6ff8fe45150c37488ebb-Abstract.html},
	urldate = {2022-12-20},
	booktitle = {Advances in {Neural} {Information} {Processing} {Systems}},
	publisher = {Curran Associates, Inc.},
	author = {Beliy, Roman and Gaziv, Guy and Hoogi, Assaf and Strappini, Francesca and Golan, Tal and Irani, Michal},
	year = {2019},
}

@article{guclu_deep_2015,
	title = {Deep {Neural} {Networks} {Reveal} a {Gradient} in the {Complexity} of {Neural} {Representations} across the {Ventral} {Stream}},
	volume = {35},
	copyright = {Copyright © 2015 the authors 0270-6474/15/3510005-10\$15.00/0},
	issn = {0270-6474, 1529-2401},
	url = {https://www.jneurosci.org/content/35/27/10005},
	doi = {10.1523/JNEUROSCI.5023-14.2015},
	language = {en},
	number = {27},
	urldate = {2022-12-20},
	journal = {Journal of Neuroscience},
	author = {Güçlü, Umut and Gerven, Marcel A. J. van},
	month = jul,
	year = {2015},
	pmid = {26157000},
	note = {Publisher: Society for Neuroscience
Section: Articles},
	pages = {10005--10014},
}

@article{olah_feature_2017,
	title = {Feature {Visualization}},
	volume = {2},
	issn = {2476-0757},
	url = {https://distill.pub/2017/feature-visualization},
	doi = {10.23915/distill.00007},
	language = {en},
	number = {11},
	urldate = {2022-12-20},
	journal = {Distill},
	author = {Olah, Chris and Mordvintsev, Alexander and Schubert, Ludwig},
	month = nov,
	year = {2017},
	pages = {e7},
}

@inproceedings{krizhevsky_imagenet_2012,
	title = {{ImageNet} {Classification} with {Deep} {Convolutional} {Neural} {Networks}},
	volume = {25},
	url = {https://proceedings.neurips.cc/paper/2012/hash/c399862d3b9d6b76c8436e924a68c45b-Abstract.html},
	urldate = {2022-12-20},
	booktitle = {Advances in {Neural} {Information} {Processing} {Systems}},
	publisher = {Curran Associates, Inc.},
	author = {Krizhevsky, Alex and Sutskever, Ilya and Hinton, Geoffrey E},
	year = {2012},
}

@article{st-yves_brain-optimized_2023,
	title = {Brain-optimized deep neural network models of human visual areas learn non-hierarchical representations},
	volume = {14},
	copyright = {2023 This is a U.S. Government work and not under copyright protection in the US; foreign copyright protection may apply},
	issn = {2041-1723},
	url = {https://www.nature.com/articles/s41467-023-38674-4},
	doi = {10.1038/s41467-023-38674-4},
	language = {en},
	number = {1},
	urldate = {2024-01-02},
	journal = {Nature Communications},
	author = {St-Yves, Ghislain and Allen, Emily J. and Wu, Yihan and Kay, Kendrick and Naselaris, Thomas},
	month = jun,
	year = {2023},
	note = {Number: 1
Publisher: Nature Publishing Group},
	pages = {3329},
}

@article{wang_better_2023,
	title = {Better models of human high-level visual cortex emerge from natural language supervision with a large and diverse dataset},
	volume = {5},
	copyright = {2023 The Author(s), under exclusive licence to Springer Nature Limited},
	issn = {2522-5839},
	url = {https://www.nature.com/articles/s42256-023-00753-y},
	doi = {10.1038/s42256-023-00753-y},
	language = {en},
	number = {12},
	urldate = {2024-01-02},
	journal = {Nature Machine Intelligence},
	author = {Wang, Aria Y. and Kay, Kendrick and Naselaris, Thomas and Tarr, Michael J. and Wehbe, Leila},
	month = dec,
	year = {2023},
	note = {Number: 12
Publisher: Nature Publishing Group},
	pages = {1415--1426},
}

@article{kay_identifying_2008,
	title = {Identifying natural images from human brain activity},
	volume = {452},
	copyright = {2008 Springer Nature Limited},
	issn = {1476-4687},
	url = {https://www.nature.com/articles/nature06713},
	doi = {10.1038/nature06713},
	language = {en},
	number = {7185},
	urldate = {2024-01-04},
	journal = {Nature},
	author = {Kay, Kendrick N. and Naselaris, Thomas and Prenger, Ryan J. and Gallant, Jack L.},
	month = mar,
	year = {2008},
	note = {Number: 7185
Publisher: Nature Publishing Group},
	pages = {352--355},
}

@article{shen_deep_2019,
	title = {Deep image reconstruction from human brain activity},
	volume = {15},
	issn = {1553-7358},
	url = {https://dx.plos.org/10.1371/journal.pcbi.1006633},
	doi = {10.1371/journal.pcbi.1006633},
	language = {en},
	number = {1},
	urldate = {2024-01-04},
	journal = {PLOS Computational Biology},
	author = {Shen, Guohua and Horikawa, Tomoyasu and Majima, Kei and Kamitani, Yukiyasu},
	editor = {O'Reilly, Jill},
	month = jan,
	year = {2019},
	pages = {e1006633},
}

@article{du_fmri_2022,
	title = {{fMRI} {Brain} {Decoding} and {Its} {Applications} in {Brain}–{Computer} {Interface}: {A} {Survey}},
	volume = {12},
	issn = {2076-3425},
	shorttitle = {{fMRI} {Brain} {Decoding} and {Its} {Applications} in {Brain}–{Computer} {Interface}},
	url = {https://www.ncbi.nlm.nih.gov/pmc/articles/PMC8869956/},
	doi = {10.3390/brainsci12020228},
	number = {2},
	urldate = {2024-01-04},
	journal = {Brain Sciences},
	author = {Du, Bing and Cheng, Xiaomu and Duan, Yiping and Ning, Huansheng},
	month = feb,
	year = {2022},
	pmid = {35203991},
	pmcid = {PMC8869956},
	pages = {228},
}

@article{rakhimberdina_natural_2021,
	title = {Natural {Image} {Reconstruction} {From} {fMRI} {Using} {Deep} {Learning}: {A} {Survey}},
	volume = {15},
	issn = {1662-453X},
	shorttitle = {Natural {Image} {Reconstruction} {From} {fMRI} {Using} {Deep} {Learning}},
	url = {https://www.frontiersin.org/articles/10.3389/fnins.2021.795488},
	urldate = {2024-01-04},
	journal = {Frontiers in Neuroscience},
	author = {Rakhimberdina, Zarina and Jodelet, Quentin and Liu, Xin and Murata, Tsuyoshi},
	year = {2021},
}

@inproceedings{st-yves_generative_2018,
	title = {Generative {Adversarial} {Networks} {Conditioned} on {Brain} {Activity} {Reconstruct} {Seen} {Images}},
	url = {https://ieeexplore.ieee.org/document/8616183},
	doi = {10.1109/SMC.2018.00187},
	urldate = {2024-01-04},
	booktitle = {2018 {IEEE} {International} {Conference} on {Systems}, {Man}, and {Cybernetics} ({SMC})},
	author = {St-Yves, Ghislain and Naselaris, Thomas},
	month = oct,
	year = {2018},
	note = {ISSN: 2577-1655},
	pages = {1054--1061},
}

@article{seeliger_generative_2018,
	title = {Generative adversarial networks for reconstructing natural images from brain activity},
	volume = {181},
	issn = {1053-8119},
	url = {https://www.sciencedirect.com/science/article/pii/S105381191830658X},
	doi = {10.1016/j.neuroimage.2018.07.043},
	urldate = {2024-01-04},
	journal = {NeuroImage},
	author = {Seeliger, K. and Güçlü, U. and Ambrogioni, L. and Güçlütürk, Y. and van Gerven, M. A. J.},
	month = nov,
	year = {2018},
	pages = {775--785},
}

@article{du_reconstructing_2019,
	title = {Reconstructing {Perceived} {Images} {From} {Human} {Brain} {Activities} {With} {Bayesian} {Deep} {Multiview} {Learning}},
	volume = {30},
	issn = {2162-2388},
	url = {https://ieeexplore.ieee.org/abstract/document/8574054?casa_token=uIVSkvSBtpUAAAAA:hLSXAWUWLM1VDzNja8psuGNxSWmo33Xu55-b4inB9gXEjSLEo1GaF7muYTfEDXwVIxY-Ltu1},
	doi = {10.1109/TNNLS.2018.2882456},
	number = {8},
	urldate = {2024-01-04},
	journal = {IEEE Transactions on Neural Networks and Learning Systems},
	author = {Du, Changde and Du, Changying and Huang, Lijie and He, Huiguang},
	month = aug,
	year = {2019},
	note = {Conference Name: IEEE Transactions on Neural Networks and Learning Systems},
	pages = {2310--2323},
}

@misc{takagi_high-resolution_2023,
	title = {High-resolution image reconstruction with latent diffusion models from human brain activity},
	copyright = {© 2023, Posted by Cold Spring Harbor Laboratory. This pre-print is available under a Creative Commons License (Attribution 4.0 International), CC BY 4.0, as described at http://creativecommons.org/licenses/by/4.0/},
	url = {https://www.biorxiv.org/content/10.1101/2022.11.18.517004v3},
	doi = {10.1101/2022.11.18.517004},
	language = {en},
	urldate = {2024-01-04},
	publisher = {bioRxiv},
	author = {Takagi, Yu and Nishimoto, Shinji},
	month = mar,
	year = {2023},
	note = {Pages: 2022.11.18.517004
Section: New Results},
}

@misc{ozcelik_natural_2023,
	title = {Natural scene reconstruction from {fMRI} signals using generative latent diffusion},
	url = {http://arxiv.org/abs/2303.05334},
	language = {en},
	urldate = {2024-01-04},
	publisher = {arXiv},
	author = {Ozcelik, Furkan and VanRullen, Rufin},
	month = jun,
	year = {2023},
	note = {arXiv:2303.05334 [cs, q-bio]},
}

@misc{dhariwal_diffusion_2021,
	title = {Diffusion {Models} {Beat} {GANs} on {Image} {Synthesis}},
	url = {http://arxiv.org/abs/2105.05233},
	urldate = {2024-01-04},
	publisher = {arXiv},
	author = {Dhariwal, Prafulla and Nichol, Alex},
	month = jun,
	year = {2021},
	note = {arXiv:2105.05233 [cs, stat]},
}

@misc{sundararajan_axiomatic_2017,
	title = {Axiomatic {Attribution} for {Deep} {Networks}},
	url = {http://arxiv.org/abs/1703.01365},
	urldate = {2024-01-05},
	publisher = {arXiv},
	author = {Sundararajan, Mukund and Taly, Ankur and Yan, Qiqi},
	month = jun,
	year = {2017},
	note = {arXiv:1703.01365 [cs]},
}

@article{walker_inception_2019,
	title = {Inception loops discover what excites neurons most using deep predictive models},
	volume = {22},
	issn = {1097-6256, 1546-1726},
	url = {https://www.nature.com/articles/s41593-019-0517-x},
	doi = {10.1038/s41593-019-0517-x},
	language = {en},
	number = {12},
	urldate = {2024-01-08},
	journal = {Nature Neuroscience},
	author = {Walker, Edgar Y. and Sinz, Fabian H. and Cobos, Erick and Muhammad, Taliah and Froudarakis, Emmanouil and Fahey, Paul G. and Ecker, Alexander S. and Reimer, Jacob and Pitkow, Xaq and Tolias, Andreas S.},
	month = dec,
	year = {2019},
	pages = {2060--2065},
}

@article{gu_neurogen_2022,
	title = {{NeuroGen}: {Activation} optimized image synthesis for discovery neuroscience},
	volume = {247},
	issn = {1053-8119},
	shorttitle = {{NeuroGen}},
	url = {https://www.sciencedirect.com/science/article/pii/S1053811921010831},
	doi = {10.1016/j.neuroimage.2021.118812},
	urldate = {2024-01-08},
	journal = {NeuroImage},
	author = {Gu, Zijin and Jamison, Keith Wakefield and Khosla, Meenakshi and Allen, Emily J. and Wu, Yihan and St-Yves, Ghislain and Naselaris, Thomas and Kay, Kendrick and Sabuncu, Mert R. and Kuceyeski, Amy},
	month = feb,
	year = {2022},
	pages = {118812},
}

@article{gu_human_2023,
	title = {Human brain responses are modulated when exposed to optimized natural images or synthetically generated images},
	volume = {6},
	copyright = {2023 The Author(s)},
	issn = {2399-3642},
	url = {https://www.nature.com/articles/s42003-023-05440-7},
	doi = {10.1038/s42003-023-05440-7},
	language = {en},
	number = {1},
	urldate = {2024-01-08},
	journal = {Communications Biology},
	author = {Gu, Zijin and Jamison, Keith and Sabuncu, Mert R. and Kuceyeski, Amy},
	month = oct,
	year = {2023},
	note = {Number: 1
Publisher: Nature Publishing Group},
	pages = {1--12},
}

@article{engel_retinotopic_1997,
	title = {Retinotopic organization in human visual cortex and the spatial precision of functional {MRI}},
	volume = {7},
	doi = {10.1093/cercor/7.2.181},
	journal = {Cerebral cortex (New York, N.Y. : 1991)},
	author = {Engel, Stephen and Glover, G.H. and Wandell, Brian},
	month = apr,
	year = {1997},
	pages = {181--92},
}

@misc{wang_disentangled_2023,
	title = {Disentangled {Representation} {Learning}},
	url = {http://arxiv.org/abs/2211.11695},
	urldate = {2024-04-02},
	publisher = {arXiv},
	author = {Wang, Xin and Chen, Hong and Tang, Si'ao and Wu, Zihao and Zhu, Wenwu},
	month = aug,
	year = {2023},
	note = {arXiv:2211.11695 [cs]},
}

@article{lecun_gradient-based_1998,
	title = {Gradient-{Based} {Learning} {Applied} to {Document} {Recognition}},
	language = {en},
	author = {LeCun, Yann and Bottou, Leon and Bengio, Yoshua and Ha, Patrick},
	year = {1998},
}

@article{agrawal_convolutional_nodate,
	title = {Convolutional {Neural} {Networks} {Mimic} the {Hierarchy} of {Visual} {Representations} in the {Human} {Brain}},
	language = {en},
	author = {Agrawal, Pulkit and Malik, Jitendra and Stansbury, Dustin and Gallant, Jack},
}

@article{hubel_sequence_1974,
	title = {Sequence regularity and geometry of orientation columns in the monkey striate cortex},
	volume = {158},
	issn = {0021-9967},
	doi = {10.1002/cne.901580304},
	language = {eng},
	number = {3},
	journal = {The Journal of Comparative Neurology},
	author = {Hubel, D. H. and Wiesel, T. N.},
	month = dec,
	year = {1974},
	pmid = {4436456},
	pages = {267--293},
}

@book{huettel_functional_2014,
	address = {Oxford, New York},
	edition = {Third Edition, Third Edition},
	title = {Functional {Magnetic} {Resonance} {Imaging}},
	isbn = {978-0-87893-627-4},
	publisher = {Oxford University Press},
	author = {Huettel, Scott A. and Song, Allen W. and McCarthy, {and} Gregory},
	month = aug,
	year = {2014},
}

@inproceedings{vaswani_attention_2017,
	title = {Attention is {All} you {Need}},
	volume = {30},
	url = {https://proceedings.neurips.cc/paper_files/paper/2017/hash/3f5ee243547dee91fbd053c1c4a845aa-Abstract.html},
	urldate = {2024-11-10},
	booktitle = {Advances in {Neural} {Information} {Processing} {Systems}},
	publisher = {Curran Associates, Inc.},
	author = {Vaswani, Ashish and Shazeer, Noam and Parmar, Niki and Uszkoreit, Jakob and Jones, Llion and Gomez, Aidan N and Kaiser, Ł ukasz and Polosukhin, Illia},
	year = {2017},
}

@misc{simonyan_very_2015,
	title = {Very {Deep} {Convolutional} {Networks} for {Large}-{Scale} {Image} {Recognition}},
	url = {http://arxiv.org/abs/1409.1556},
	doi = {10.48550/arXiv.1409.1556},
	urldate = {2024-11-10},
	publisher = {arXiv},
	author = {Simonyan, Karen and Zisserman, Andrew},
	month = apr,
	year = {2015},
	note = {arXiv:1409.1556},
}

@misc{kiat_greentfrapplucent_2024,
	title = {greentfrapp/lucent},
	copyright = {Apache-2.0},
	url = {https://github.com/greentfrapp/lucent},
	urldate = {2024-11-10},
	author = {Kiat, Lim Swee},
	month = nov,
	year = {2024},
	note = {original-date: 2020-05-09T18:07:01Z},
}

@article{jing2019neural,
  title={Neural style transfer: A review},
  author={Jing, Yongcheng and Yang, Yezhou and Feng, Zunlei and Ye, Jingwen and Yu, Yizhou and Song, Mingli},
  journal={IEEE transactions on visualization and computer graphics},
  volume={26},
  number={11},
  pages={3365--3385},
  year={2019},
  publisher={IEEE}
}

@inproceedings{radford2021learning,
  title={Learning transferable visual models from natural language supervision},
  author={Radford, Alec and Kim, Jong Wook and Hallacy, Chris and Ramesh, Aditya and Goh, Gabriel and Agarwal, Sandhini and Sastry, Girish and Askell, Amanda and Mishkin, Pamela and Clark, Jack and others},
  booktitle={International conference on machine learning},
  pages={8748--8763},
  year={2021},
  organization={PMLR}
}

@book{bird2009natural,
  title={Natural language processing with Python: analyzing text with the natural language toolkit},
  author={Bird, Steven and Klein, Ewan and Loper, Edward},
  year={2009},
  publisher={" O'Reilly Media, Inc."}
}

@article{francis1979brown,
  title={Brown corpus manual, manual of information to accompany a standard corpus of present-day edited American English},
  author={Francis, W and Kucera, Henry},
  journal={Dept. of Linguistics, Brown Univ., Tech. Rep},
  year={1979}
}

\newpage

\section{Appendix}\label{Appendix} Here we show the results for the retinotopy, dreams, and
attribution for subjects, 2, 5 and 7.
\begin{figure}[h]
    \centering
    \includegraphics[width=\linewidth]{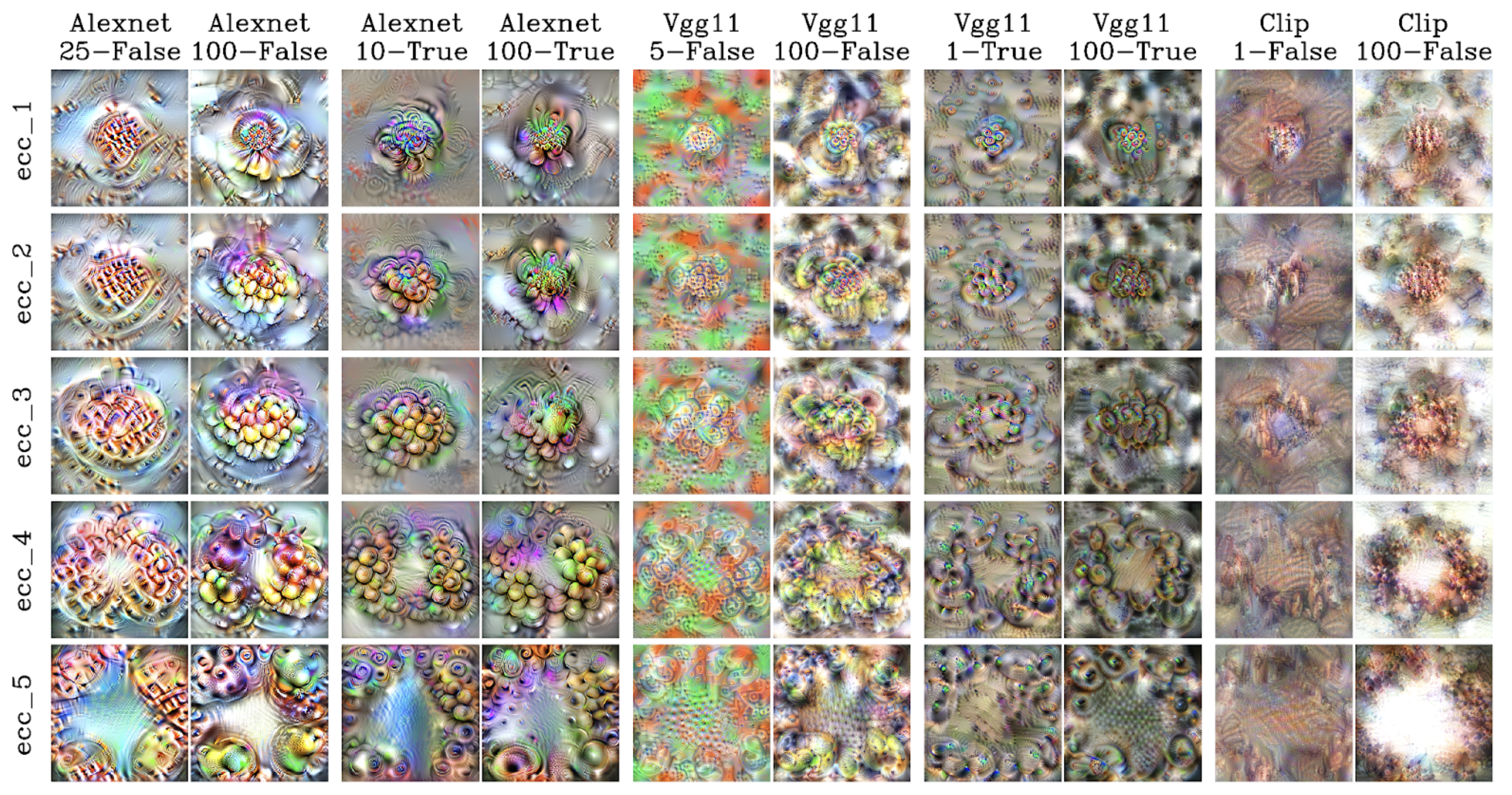}
    \caption{This figure shows the results for subject 2 of the dreams for the eccentricity ROIs (row) for the best backbones and the corresponding one with 100\% filters (columns). Each column is a backbone specification denoted in the title as <backbone>-<percent>-<fine tuning>.    }
    \label{fig:ecc_subj-2}
\end{figure}
\begin{figure}
    \centering
    \includegraphics[width=\linewidth]{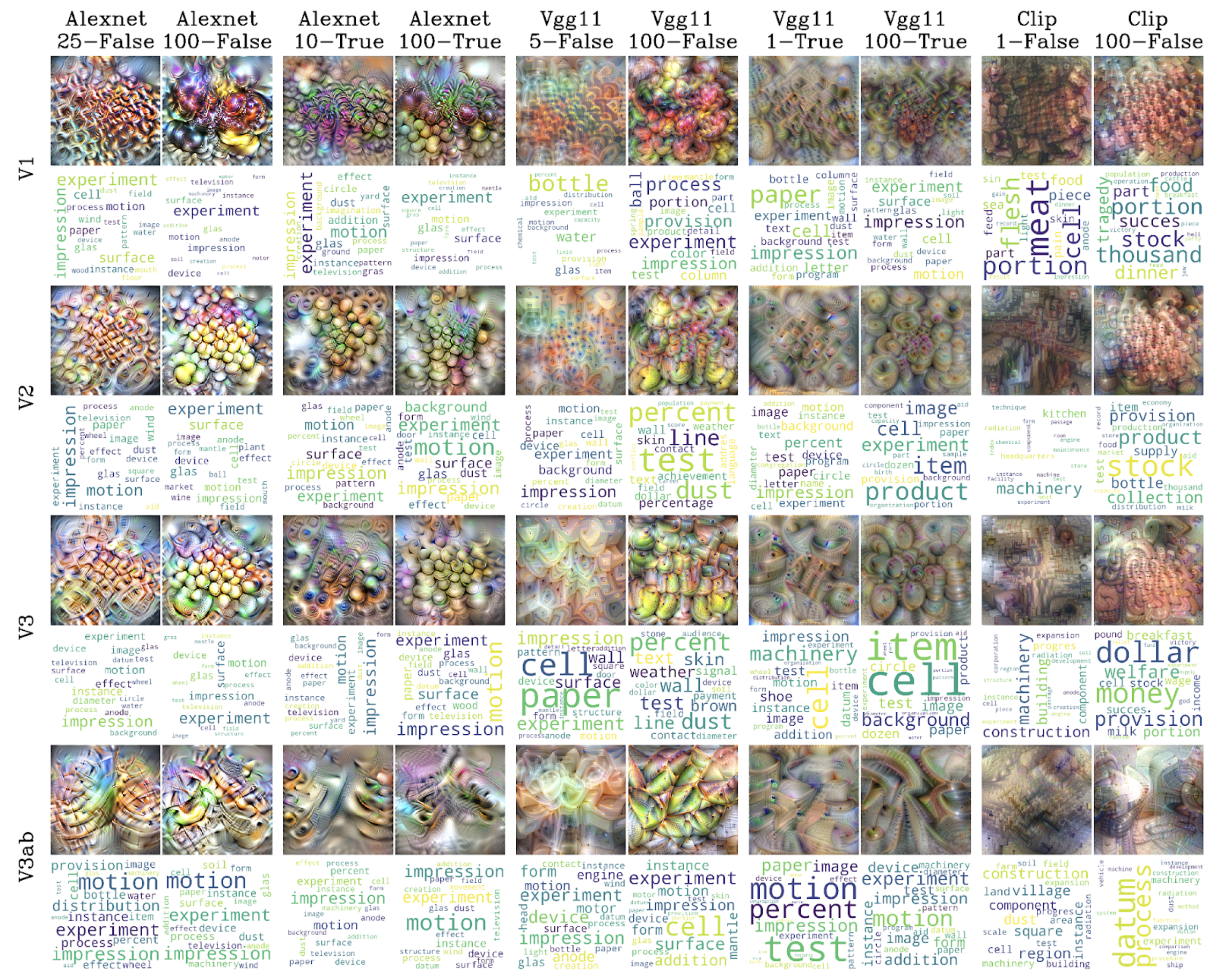}
    \caption{This figure shows the results for subject 2 of the dreams, with word clouds underneath, for the early visual cortex areas V1,V2, V3 and V3ab (rows) for the best backbones (columns). The size of the words denote the similarity with the image.}
    \label{fig:enter-label}
\end{figure}
\begin{figure}
    \centering
    \includegraphics[width=\linewidth]{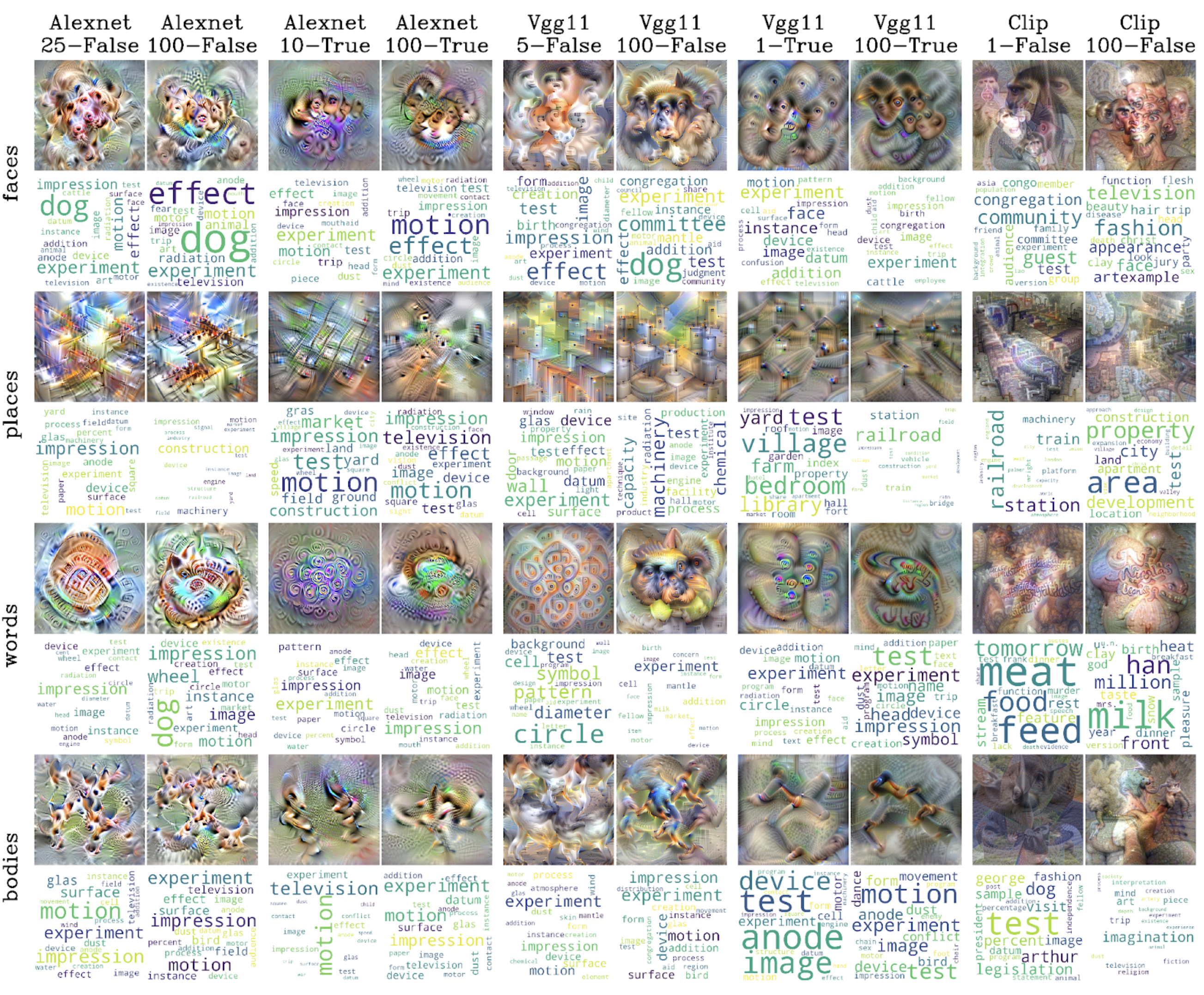}
    \caption{This figure shows the results for subject 2 of the dreams, with word clouds underneath, for the higher visual cortex areas corresponding to faces (OFA, FFA, mTL-faces and aTL-faces), places (OPA, PPA and RSC), words (OWFA, VMFA, mfs-words, and mTL-words) and bodies (EBA, FBA and mTL-bodies) for the best backbones (columns). The size of the words denote the similarity with the image.  }
    \label{fig:enter-label}
\end{figure}
\begin{figure}
    \centering
    \includegraphics[width=\linewidth]{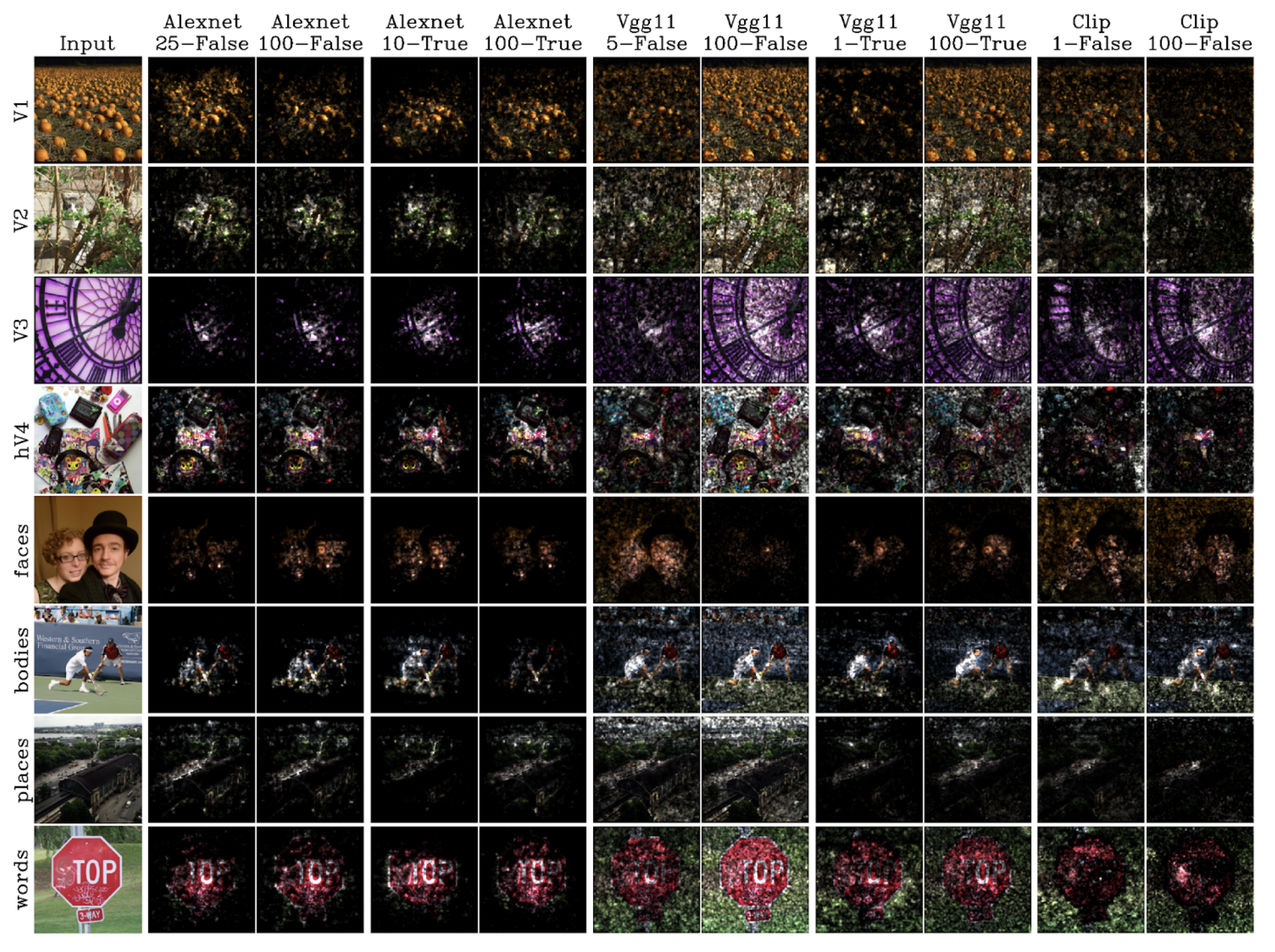}
    \caption{This figure shows, subject 2, the implicit attention from the integrated gradient approach as an intensity mask. The first column is an input image, and the remaining columns are the different backbone configurations, titled as <backbone>-<percent>-<fine tuning>. The rows show the different ROIs.   }
    \label{fig:enter-label}
\end{figure}

\begin{figure}
    \centering
    \includegraphics[width=\linewidth]{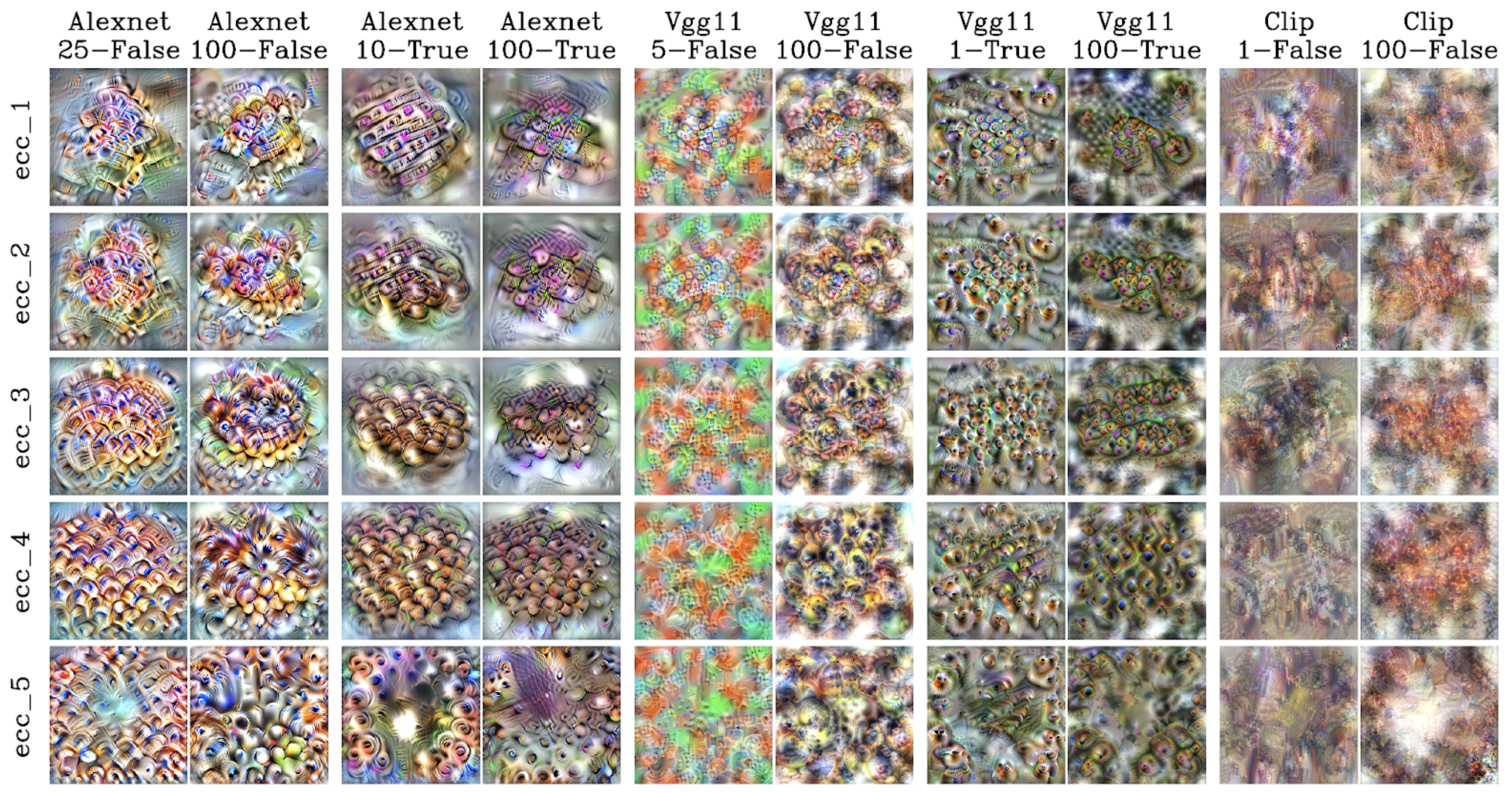}
    \caption{This figure shows the results for subject 5 of the dreams for the eccentricity ROIs (row) for the best backbones and the corresponding one with 100\% filters (columns). Each column is a backbone specification denoted in the title as <backbone>-<percent>-<fine tuning>.   }
    \label{fig:ecc_subj-5}
\end{figure}
\begin{figure}
    \centering
    \includegraphics[width=\linewidth]{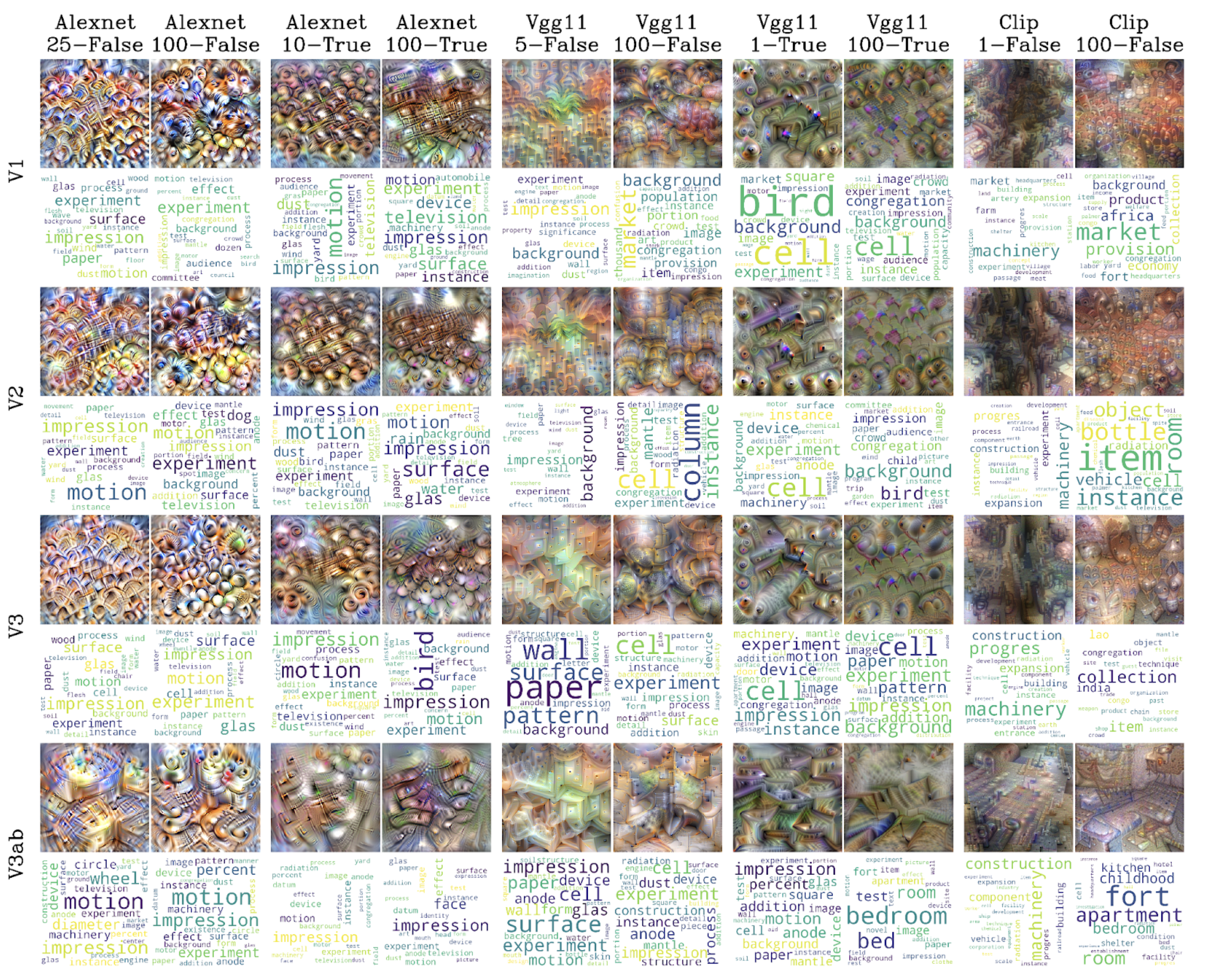}
    \caption{This figure shows the results for subject 5 of the dreams, with word clouds underneath, for the early visual cortex areas V1,V2, V3 and V3ab (rows) for the best backbones (columns). The size of the words denote the similarity with the image.}
    \label{fig:enter-label}
\end{figure}
\begin{figure}
    \centering
    \includegraphics[width=\linewidth]{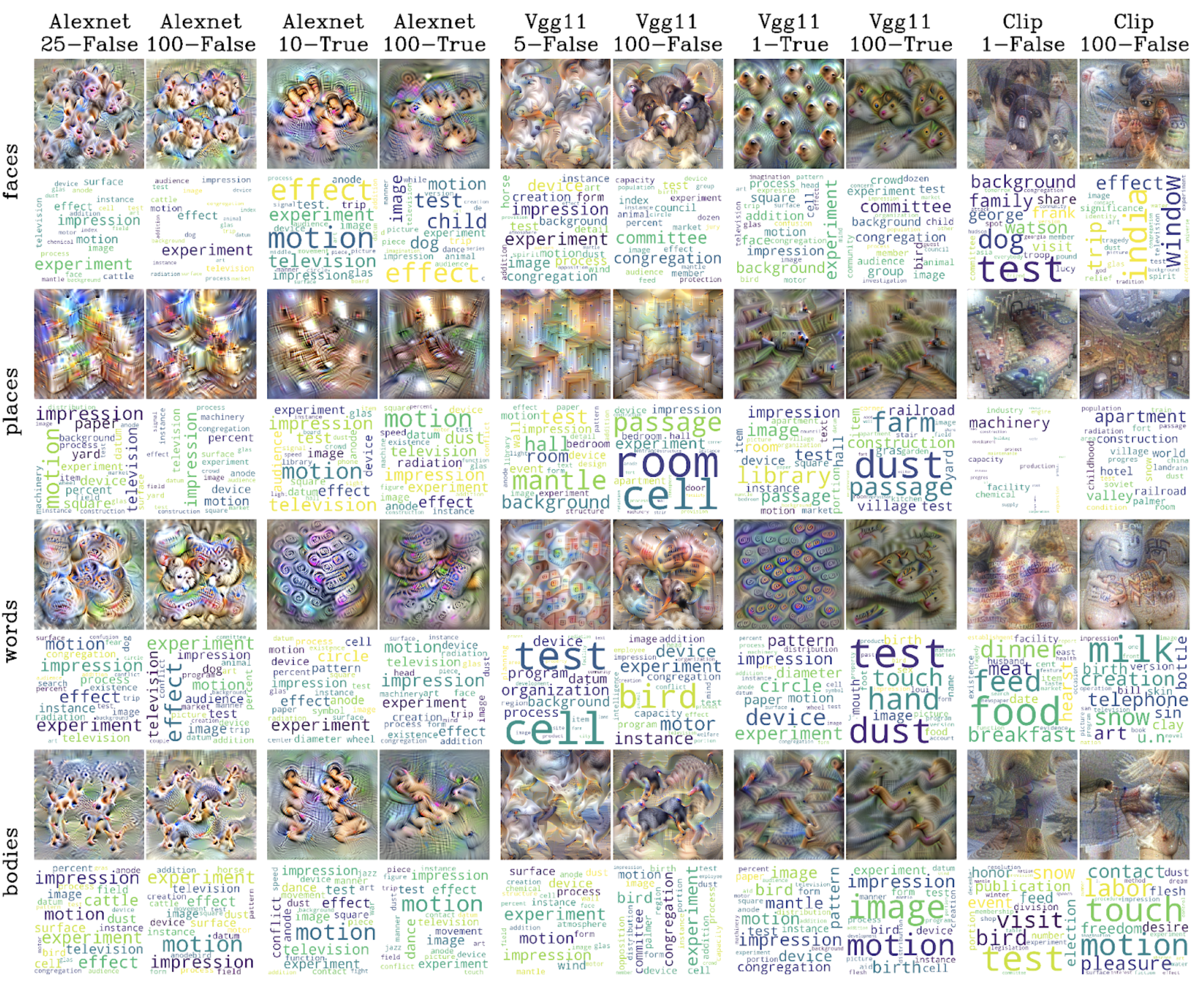}
    \caption{This figure shows the results for subject 5 of the dreams, with word clouds underneath, for the higher visual cortex areas corresponding to faces (OFA, FFA, mTL-faces and aTL-faces), places (OPA, PPA and RSC), words (OWFA, VMFA, mfs-words, and mTL-words) and bodies (EBA, FBA and mTL-bodies) for the best backbones (columns). The size of the words denote the similarity with the image.  }
    \label{fig:enter-label}
\end{figure}
\begin{figure}
    \centering
    \includegraphics[width=\linewidth]{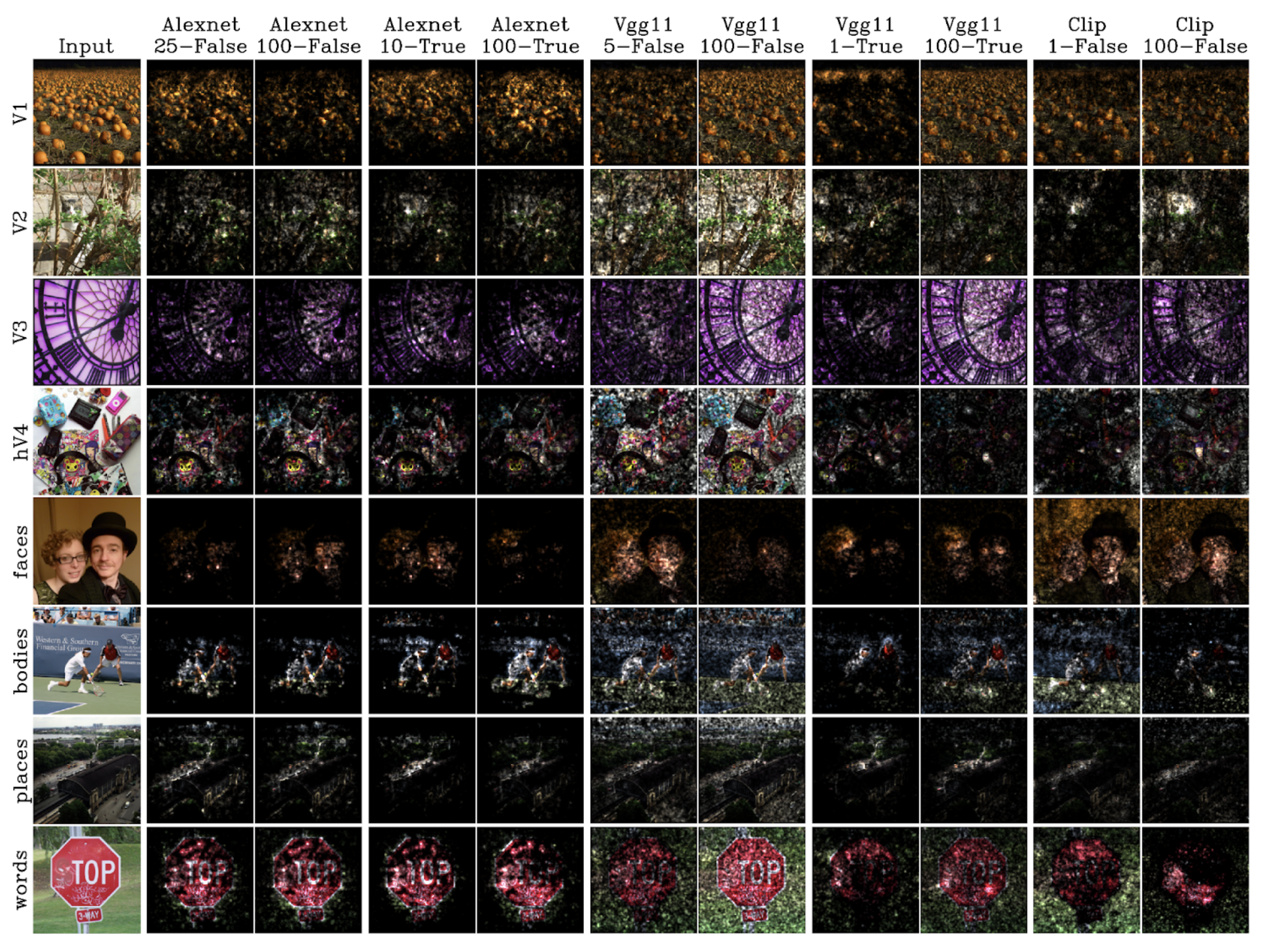}
    \caption{This figure shows, subject 5, the implicit attention from the integrated gradient approach as an intensity mask. The first column is an input image, and the remaining columns are the different backbone configurations, titled as <backbone>-<percent>-<fine tuning>. The rows show the different ROIs.   }
    \label{fig:enter-label}
\end{figure}

\begin{figure}
    \centering
    \includegraphics[width=\linewidth]{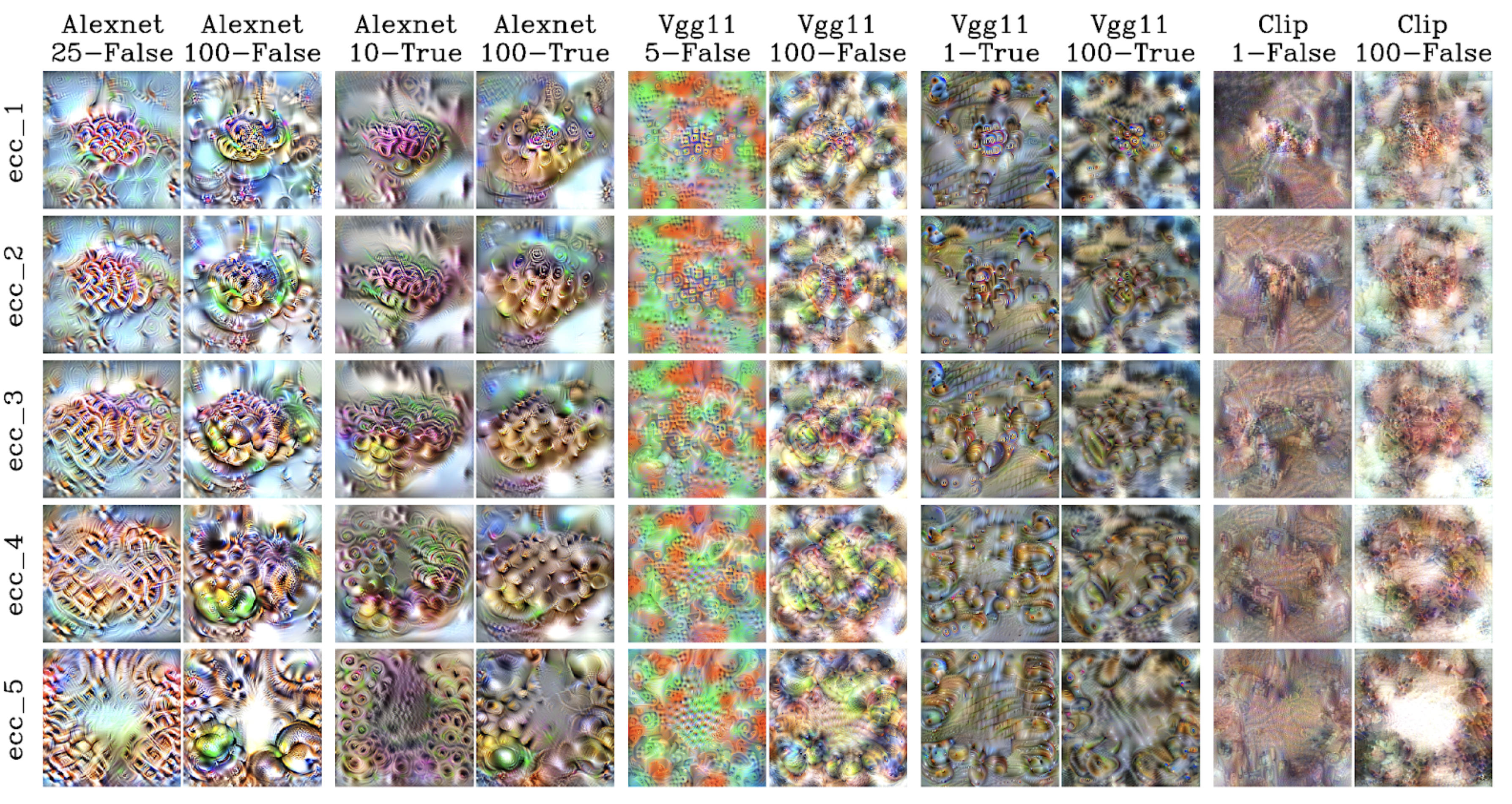}
    \caption{This figure shows the results for subject 7 of the dreams for the eccentricity ROIs (row) for the best backbones and the corresponding one with 100\% filters (columns). Each column is a backbone specification denoted in the title as <backbone>-<percent>-<fine tuning>.   }
    \label{fig:ecc_subj-7}
\end{figure}
\begin{figure}
    \centering
    \includegraphics[width=\linewidth]{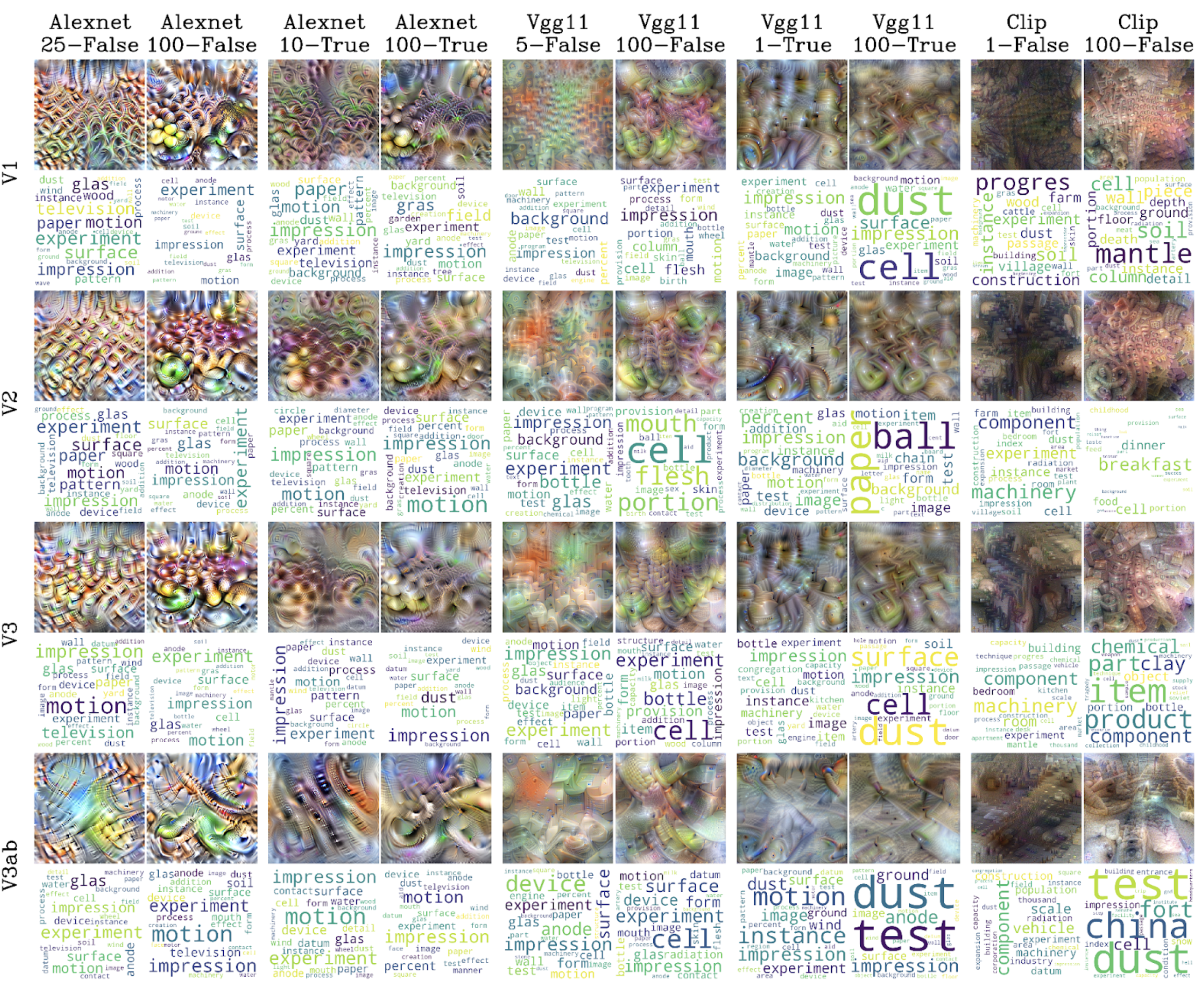}
    \caption{This figure shows the results for subject 7 of the dreams, with word clouds underneath, for the early visual cortex areas V1,V2, V3 and V3ab (rows) for the best backbones (columns). The size of the words denote the similarity with the image.}
    \label{fig:enter-label}
\end{figure}
\begin{figure}
    \centering
    \includegraphics[width=\linewidth]{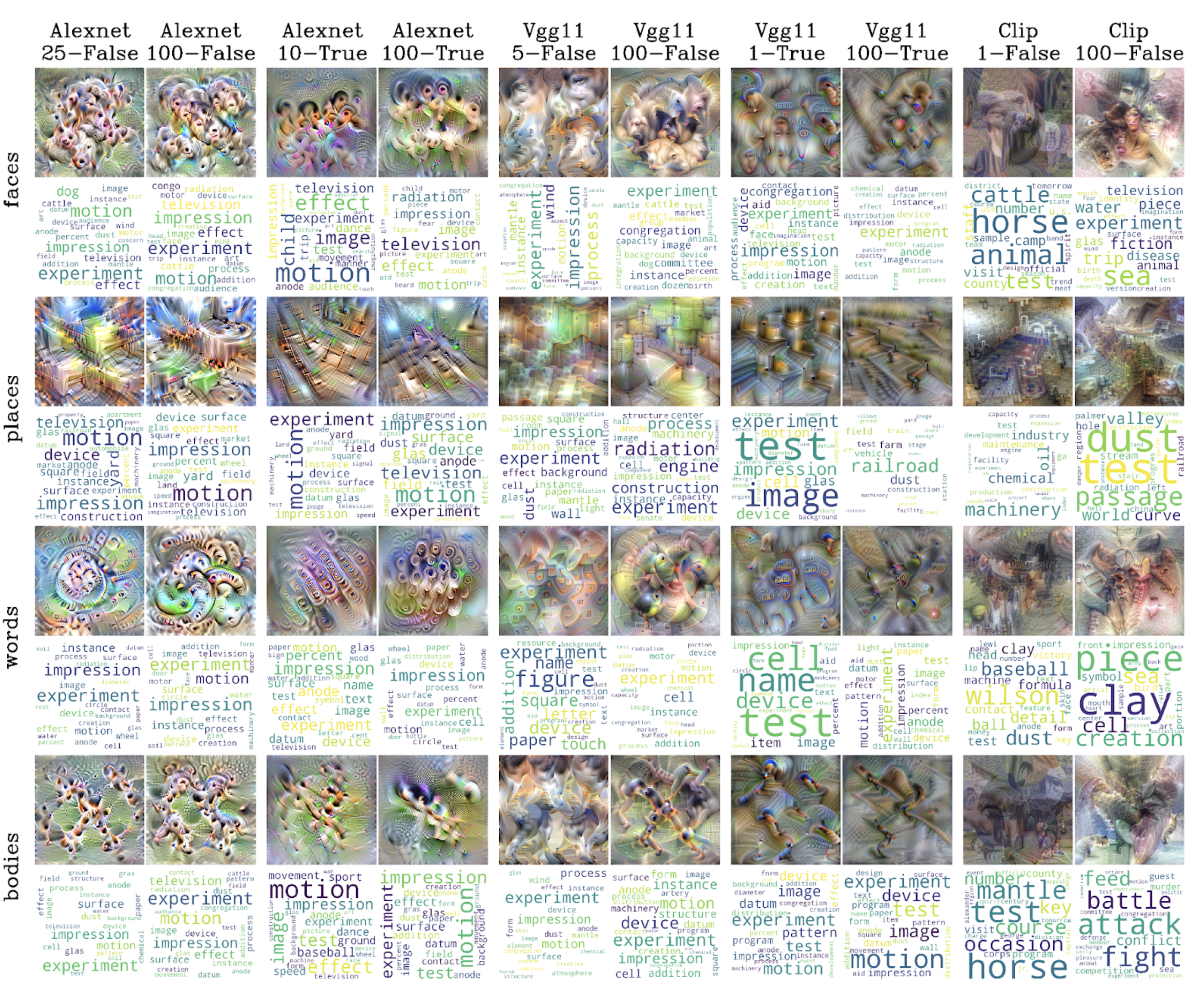}
    \caption{This figure shows the results for subject 7 of the dreams, with word clouds underneath, for the higher visual cortex areas corresponding to faces (OFA, FFA, mTL-faces and aTL-faces), places (OPA, PPA and RSC), words (OWFA, VMFA, mfs-words, and mTL-words) and bodies (EBA, FBA and mTL-bodies) for the best backbones (columns). The size of the words denote the similarity with the image.  }
    \label{fig:enter-label}
\end{figure}
\begin{figure}
    \centering
    \includegraphics[width=\linewidth]{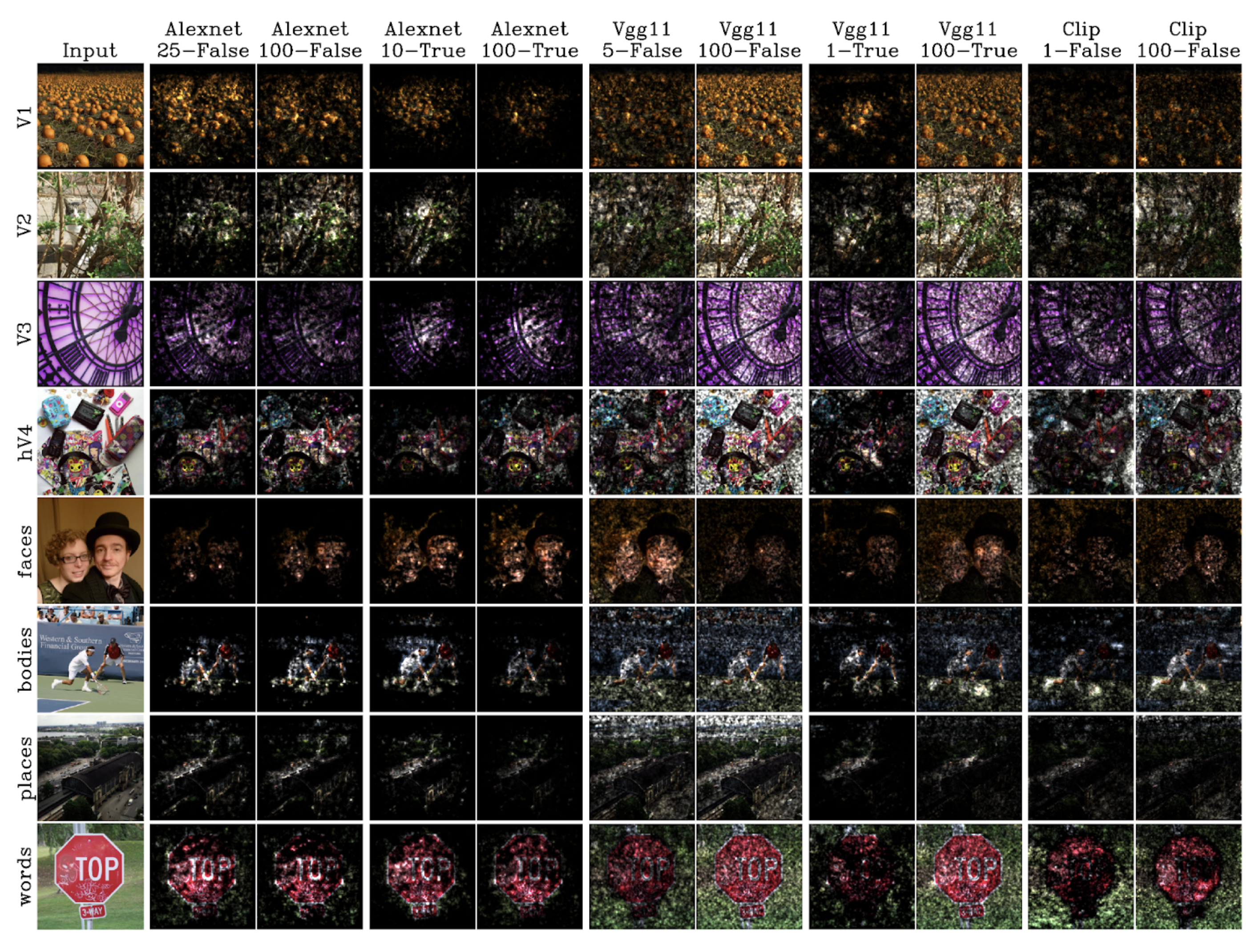}
    \caption{This figure shows, subject 7, the implicit attention from the integrated gradient approach as an intensity mask. The first column is an input image, and the remaining columns are the different backbone configurations, titled as <backbone>-<percent>-<fine tuning>. The rows show the different ROIs.   }
    \label{fig:enter-label}
\end{figure}
\end{document}